\documentstyle[12pt]{article}
\begin{document}
\begin{center} {\Large {\bf {Multiple dynamic transitions in 
anisotropic Heisenberg ferromagnet driven by
polarised magnetic field}}}\end{center}

\vskip 1 cm

\begin{center} {\bf {Muktish Acharyya}}\end{center}

\vskip 0.5 cm

\begin{center}{\it Department of Physics, Krishnanagar Government College,}\\
{\it P.O.-Krishnanagar, Dist-Nadia, PIN-741101, West-Bengal, India}\\
{\it E-mail:muktish@vsnl.net}\end{center}

\vskip 2 cm

\noindent {\bf {Abstract:}} Uniaxially (along Z-axis) anisotropic Heisenberg ferromagnet, in presence of time dependent
(but uniform over space)
magnetic field, is studied by Monte Carlo simulation. The time dependent magnetic field was taken as
elliptically polarised where the resultant field vector rotates in X-Z plane. The system is cooled 
(in presence of the elliptically polarised magnetic field) from high
temperature. As the temperature decreases, it was found that in the low anisotropy limit 
the system undergoes three successive dynamical phase transitions.
These three dynamic transitions were confirmed by studying the temperature variation of dynamic 'specific heat'.
The temperature variation of dynamic 'specific heat' shows three peaks indicating three dynamic transition 
points. 

\vskip 0.5 cm


\vskip 1.5 cm
\noindent {\bf Introduction:}

The dynamical behaviours of magnetic model systems, in the presence of time dependent magnetic field, 
show interesting physical phenomena \cite{rev}. The nonequilibrium dynamical
phase transition \cite{rev}, particularly in the kinetic Ising model has drawn much interest of researchers in
the field of nonequilibrium statistical physics. The dynamic transition in the kinetic Ising model in the 
presence of magnetic field (sinusoidally varying in time) was first noticed \cite{tom} in the mean field
solution of the dynamical equation for the average magnetisation. The time averaged magnetisation over a full
cycle (of external magnetic field) becomes nonzero at finite values of temperature and field amplitude.
These values of temperature and field amplitudes depend on the frequency of oscillating field. However, the
transition there \cite{tom} is not perfectly dynamic in nature since it can exist for such equation of motion
even in static (zero frequency) limit ! This reveals that the transition in the zero frequency limit is an
artefact of the meanfield approximation which does not consider the nontrivial fluctuations. The occurence
of the true dynamic transitions for models, incorporating the thermodynamical fluctuations, was later shown in
several Monte Carlo studies \cite{rev}.

After observing the true dynamic transition in the kinetic Ising model in presence of oscillating magnetic
field and knowing that it is a nonequilibrium transition, a considerable amount of studies were performed \cite{rev} to
establish that this transition is thermodynamic phase transition. The divergences of 'time scale' \cite{ma1} and the
'dynamic specific heat' \cite{ma1} and the divergence of length scale \cite{sides} are two important observations
to establish that the dynamic transition is a thermodynamic phase transition. 

Although the dynamic transition in kinetic Ising model is an interesting phenomenon and a simple example to
grasp the various features of nonequilibrium phase transitions, it has several limitations. In the Ising model,
since the spins can have only two orientations (up/down), some interesting features of dynamic transitions 
(related to the dynamic transverse ordering) are missing
in this model. The classical vector spin model\cite{book} would be the better choice to see such interesting phenomena which
are missing in Ising model. One of such examples is the 'off-axial' dynamic transitions \cite{ijmpc} recently observed in
the anisotropic Heisenberg ferromagnet. In the 'off-axial' dynamic transition, the dynamical symmetry along the axis of 
anisotropy can be broken by applying an oscillating field along any perpendicular direction (X-direction say). 
The dynamic phase transition
in anisotropic XY spin system in an oscillating magnetic field is recently studied \cite{yasui} by solving Ginzburg-Landau
equation. The dynamic phase transition and the dependence of its behaviour on the bilinear exchange anisotropy of a
classical Heisenberg spin system (planar thin ferromagnetic film), is recently studied \cite{jang} by Monte Carlo simulation.

All these studies on the dynamic phase transition, made so far, are related to single transition. The dynamic transition
occurs at a single value of temperature (for fixed values of field amplitude and frequency). 
No evidence of multiple dynamic transitions (for bulk only) is reported so far in the
literature in anisotropic Heisenberg ferromagnet driven by polarised magnetic field. However, it should be mentioned here,
that a very recent study \cite{hall} of dynamical phase transitions in thin Heisenberg ferromagnetic films with bilinear
exchange anisotropy, has shown multiple phase transitions for the surface and bulk layers of the film at different temperatures. 
Here, in this paper, the observations of multiple (triple) dynamic transitions, in anisotropic Heisenberg ferromagnet
driven by elliptically polarised magnetic field, are briefly reported 
which is observed in the
uniaxially anisotropic Heisenberg ferromagnet (three dimensional) in the presence of a elliptically polarised magnetic 
field studied by Monte Carlo simulations. 

The paper is organised as follows: the model is introduced and Monte Carlo simulation technique are
described in the next
section, 
third section contains numerical results with figures and the paper ends with a summary and
few concluding remarks given in the fourth section.

\vskip 1 cm

\noindent {\bf The model and Monte Carlo simulation technique:}

The Hamiltonian of a classical anisotropic (uniaxial and single-site) Heisenberg model \cite{book}, with nearest
neighbour ferromagnetic interaction in the presence of a magnetic field, can be represented as
\begin{equation}
H = -J \sum_{<ij>}\vec S_i \cdot \vec S_j -D \sum_i (S_i^z)^2 - \vec h \cdot \sum_i \vec S_i,
\end{equation}
\noindent where $\vec S_i [S_{ix}, S_{iy}, S_{iz}]$ represents a classical spin vector of magnitude 
unity ($S_{ix}^2+S_{iy}^2+S_{iz}^2=1)$ situated at the i-th lattice site. The classical spin vector
$\vec S_i$ can be oriented in any (unrestricted) direction in the vector spin space. In the 
above expression of the Hamiltonian,
the first term represents the nearest neighbour ($<ij>$) ferromagnetic ($J > 0$) interaction. The factor $D$
in the second term represents the strength of uniaxial (z-axis here) anisotropy which is favouring
the spin to be aligned along the z-axis. 
Here, it may be noted that for $D = 0$, the system is in the isotropic Heisenberg limit and for $D \to \infty$
the system goes to Ising limit.
The last term stands for the interaction with the externally
applied time dependent magnetic field ($\vec h [h_x,h_y,h_z] $). The magnetic field components are sinusoidally
oscillating in time i.e., $h_{\alpha} = h_{0\alpha}{\rm cos}{\omega t}$, where $h_{0\alpha}$ is the amplitude
and $\omega$ is the angular frequency ($\omega = 2\pi f; f$ is
the frequency) of the $\alpha$-th component of the magnetic field. In this present case, the field is
taken elliptically polarised. The polarised field can be represented as
\begin{equation}
\vec h = \hat x h_x + \hat y h_y + \hat z h_z\\
= \hat x h_{0x}{\rm cos}{(\omega t)}+ \hat z h_{0z}{\rm sin}{(\omega t)}.
\end{equation}
One can readily check that $h_x = h_{0x}{\rm cos}{(\omega t)}$ and $h_z = h_{0z}{\rm sin}{(\omega t)}$ yield, after the elimination 
of time, 
\begin{equation}
{{h_x^2} \over {h_{0x}^2}} + {{h_z^2} \over {h_{0z}^2}} = 1,
\end{equation}
which shows that the magnetic field lies on X-Z plane and is elliptically polarised (in general $h_{0x}$ and
$h_{0z}$ are not equal). If $h_{0x}=h_{0z}=h_0$(say), the above equation will take the form $h_x^2 + h_z^2=h_0^2$ 
and the field will be called circularly polarized. 
The magnetic fields and the strength of anisotropy ($D$) are measured in the unit
of $J$. The model is defined on a simple cubic lattice of linear size $L$ with periodic boundary conditions
applied in all three directions.

The model, described above, has been studied by Monte Carlo simulation using the following algorithm \cite{mc}.
To obtain the equilibrium spin configuration at a particular temperature $T$, the system is slowly cooled down
from a random initial spin configuration \cite{uli}. At any fixed temperature $T$ (measured in the unit of $J/K_B$, where
$K_B$ is the Boltzmann constant), and for the fixed values of $h_{0x}$, $h_{0z}$ $\omega$ and $D$, a lattice site $i$
has been chosen randomly (random updating scheme). 
Monte Carlo simulations were performed using Metropolis algorithm \cite{mc} with random updating scheme. The spin-tilt trial
configuration is generated as follows \cite{uli,ijmpc}:
               two different
               random numbers $r_1$ and $r_2$ (uniformly distributed between -1 and +1), are chosen in
               such a way that $R^2 = r_1^2 + r_2^2$ becomes less that or equal to unity. The set of
               values of $r_1$ and $r_2$, for which $R^2 > 1$, are rejected. Now, $u = {\sqrt{1-R^2}}$
               $S_{ix}=2ur_1$, $S_{iy}=2ur_2$ and $S_{iz}=1-2R^2$. In this way, the distribution of
points of tips of spin vectors on the surface of a unit sphere will be uniform. The acceptance of a trial configuration is determined
by Metropolis rate \cite{mc}. 
$L^3$ numbers of such updates 
(at random positions) of spin vectors, defines one Monte Carlo step
per site (MCSS) and this may be considered as the unit of time in this simulation. The linear frequency ($f = \omega/{2 \pi}$) of the
time varying magnetic field is taken 0.02 and kept constant throughout this simulational study. Thus 50 MCSS are required to obtain
one complete cycle of the oscillating field. Consequently, 50 MCSS is the time period ($\tau$) of the oscillating magnetic field.
Any macroscopic quantity, such as, any component of magnetisation at any instant, is calculated as follows: Starting with an initial
random spin configuration (high temperature phase), the system is allowed to become stabilised (dynamically) upto 4$\times 10^4$ MCSS
(i.e. 800 complete cycles of the oscillating field). The average value of various physical quantities are calculated from further
$4\times 10^4$ MCSS (i.e., averaged over another 800 cycles). This is important to achieve stable value and it was checked 
carefully that
the number of MCSS mentioned above is sufficient to obtain stable value of the measurable quantities etc. which can clearly
show the dynamic transitions points within limited accuracy.  
But to describe the critical behaviours very precisely (e.g., to estimate critical exponent etc.) much longer run is
necessary. Here, the total length of simulation for one fixed
temperature is $8\times 10^4$. The system is slowly cooled down ($T$ has been reduced by small interval) to obtain the values of
the statistical quantities in the low temperature ordered phase. The last spin configuration obtained at previous temperature is
used as the initial configuration for the new temperature. The CPU time required for $8\times 10^4$ MCSS is approximately 25 Minutes
on an Intel Pentium-III processor.
\vskip 1 cm

\noindent {\bf Numerical results:}

The simulational study is done for a simple cubic lattice of linear size $L = 20$. The instantaneous magnetisation
components (per lattice site) 
$m_x(t) = \sum_i {{S^i_x} \over {L^3}}$, 
$m_y(t) = \sum_i {{S^i_y} \over {L^3}}$ and
$m_z(t) = \sum_i {{S^i_z} \over {L^3}}$ are calculated at each time in the presence of magnetic field.
The time averaged (over a full cycle of the oscillating magnetic field) magnetisation components (the dynamic order
parameter components)
$Q_x = {1 \over {\tau}} \oint m_x(t) dt$,
 $Q_y = {1 \over {\tau}} \oint m_y(t) dt$ and
$Q_z = {1 \over {\tau}} \oint m_z(t) dt$ are calculated by integrating (over the complete cycle of the oscillating
field) the instantaneous magnetisation components. The total (vector) dynamic order parameter can be expressed as
$\vec Q = iQ_x + jQ_y + kQ_z$. The instantaneous energy 
$e(t) = -J \sum_{<ij>}\vec S_i \cdot \vec S_j - D \sum_i (S_i^z)^2 - \vec h \cdot \sum_i \vec S_i$ is also calculated.
The time averaged instantaneous energy is $E = {1 \over {\tau}} \oint e(t) dt$. The rate of change of $E$ with respect to
the temperature $T$ is defined as dynamic specific heat $C ( = {{dE} \over {dT}})$ \cite{ma1}. The dynamic specific heat $C$
is calculated from energy $E$, just by calculating the derivative using the three-point central difference 
formula, given below.

\begin{equation}
C = {{dE} \over {dT}} = {{E(T+{\delta T})-E(T-{\delta T})} \over {2\delta T}}
\end{equation}
For the elliptically polarised (equation 3) magnetic field, where the resultant field lies in X-Z plane, the amplitudes
of fields are taken as $h_{0x} = 0.3$ and $h_{0z} = 1.0$. The strength of uniaxial anisotropy is taken $D = 0.2$.
This value of $D$ is obtained by rigorous searching to have these interesting results and kept constant throughout
the study. However, there must be variations in transition points depending on the values of $D$. 
It is observed that for Higher values
of $D$ the multiple transition phenomenon disappears. The values of field amplitudes and frequency are also obtained
by searching.

The temperature variations of the dynamic order parameter components ($Q_x, Q_y, Q_z$) are studied and the results
are depicted in Fig.1(a). As the system is cooled down, from a high temperature disordered ($\vec Q = 0$) phase, 
it was observed that, first the system undergoes a transition
from dynamically disordered ($\vec Q$ = 0) to a dynamically {\it Y-ordered} (only $Q_y \ne 0$) phase. 
This may be called as the first phase ($P_1$) and the transition temperature is $T_{c1}$.
This phase can be characterised as $P_1$: ($Q_x = 0$, $Q_y \ne 0$, $Q_z = 0$).
Here, the resultant vector of elliptically polarised magnetic field
lies in x-z plane and the dynamic ordering occurs along y-direction. So, this is clearly an off-axial transition \cite{ijmpc}.
In the case of this type of off-axial transition the dynamical symmetry (in any direction; y-direction here) 
is broken by the application of
the magnetic field in the perpendicular direction (lies in the x-z plane here).
As the system cools down, it retains this particular dynamically ordered phase ($P_1$)
 over a considerable range of temperatures. As the temperature
decreases further, a second transion was observed. Here, the system becomes dynamically ordered both in X- and Z-directions at the
cost of Y-ordering. In this new dynamic phase,$P_2$: ($Q_x \ne 0$, $Q_y = 0$, $Q_z \ne 0$). 
In this phase the dynamical ordering is planar (lies on x-z plane). 
The ordering occurs in the same plane on which the field vector lies.
This transition is axial \cite{ijmpc}. This phase may be called the second
phase ($P_2$) and the transition (from first phase to the second phase) temperature is $T_{c2}$. 
As the temperature decreases further, the X- and Z-ordering increases.
At some lower temperature, a third transition was observed, 
from where the X-ordering starts to decrease and only Z-ordering starts to increase quite rapidly.
This third phase can be designated as $P_3$: ($Q_x \ne  0$, $Q_y = 0$, $Q_z \ne 0$).
Although the characterisation of $P_2$ and $P_3$, in terms of the values of dynamic order parameter components, looks similar 
there exists an important difference between these two phases. In the phase $P_2$, both $Q_x$ and $Q_z$ increases as the tempereture
decreases but in the phase $P_3$, $Q_x$ decreases as the temperature decreases (see Fig. 1(a)). So these two phases 
$P_2$ and $P_3$ distinctly
differs from each other.
In this phase the dynamical ordering is also axial (along Z-axis or anisotropy axis).
The system continues to increase the dynamical Z-ordering as the temperature decreases further. 
The low temperature phase is only dynamically Z-ordered. That means the systems orders dynamically (only $Q_z \ne 0$) along the
Z-direction (direction of anisotropy) only at very low temperatures. Zero temperature dynamic phase (for such polarised field)
can be characterised as $Q_x = 0$, $Q_y = 0$ and $Q_z = 1.0$.  

To detect the dynamic transitions and to find the transition temperatures the temperature variation of the energy
$E$ is plotted in Fig.1(b). From this figure it is clear that there are three dynamic transitions occur in this case.
The transition points are the inflection points in $E-T$ curve. The temperature derivative of the energy $E$ is the
dynamic specific heat $C$. The temperature variation of $C$ is shown in Fig.1(c). The three dynamic transitions
are very clearly shown by three peaks of the specific heat plotted against the temperature $T$. From this figure
the transition temperatures are calculated (from the peak positions of $C-T$ curve). First transition (right peak)
occurs around $T_{c1}=1.22$, the second transition (middle peak) occurs at $T_{c2}=0.94$ and the third (left peak) transition
occurs around $T_{c3}=0.86$.

This study was further extended for other values of $h_{0x}$ keeping other parameters fixed. It was found that this
three transitions senario disappears for higher values of $h_{0x}$. For example, for $h_{0x} = 0.9$, the second phase
$P_2$ disappears. In this case, the $C-T$ curve shows two peaks. It was also observed that for $h_{0x} = 0.2$, 
$h_{0z} = 0.2$ (keeping all other parameter fixed) the system shows single transition and only dynamically orders along
Z-direction.  

To detect the dynamic transition points an alternative method may be to study the temperature variation of fluctuation of
dynamic order parameter $\chi(Q) [ = L^3(<Q^2>-<Q>^2)]$. However, we do not have sufficient amount of precise data to
study this.

\noindent 
\vskip 1 cm

\noindent {\bf Summary:}

The uniaxially (Z-direction) anisotropic Heisenberg ferromagnet in presence of time dependent (but uniform over space) magnetic field
is studied by Monte Carlo simulation using Metropolis dynamics. The dynamic transition in uniaxially anisotropic Heisenberg 
ferromagnet is already studied by MC simulation. In that case the time dependent magnetic field was sinusoidal and the
axial and off-axial dynamic transition was reported \cite{ijmpc} earlier.

In the present study, the external time dependent magnetic field was taken elliptically polarised where the resultant
field vector rotates on X-Z plane. For the lower values of anisotropy and a specific range of the values of field
amplitudes the system undergoes multiple dynamic phase transitions. Here, three distinct phases are identified. In this paper,
this observation is just briefly reported. 
This multiple dynamic phase transition in anisotropic Heisenberg ferromagnet in presence of elliptically polarised field, is
observed here by Monte Carlo simulation. An alternative method, to check this phenomenon, may be to use Landau-Lifshitz-Gilbert
equation of motion \cite{llg} with Langevin dynamics. Another important thing should be mentioned here regarding the possible explanation
of multiple dynamic phase transitions (axial and off-axial transitions) observed in the anisotropic Heisenberg model. One
possible reason may be the coherrent rotation of spins. Where the dynamic phase transition
in the Ising model can be explained simply by spin reversal and nucleation \cite{stauffer}.
But to establish the responsible mechanism behind the multiple dynamic phase transitions, details investigations are required.

The variations of the dynamic phase boundaries with frequency and the
strength of anisotropy is quite interesting to be studied. This study also indicates that the system will show a very rich
phase diagram with multicritical behaviour. The finite size analysis is also necessary in order to distinguish the crossover
effects from the true phase transitions. This requires huge computational task which will take much time. This work is in progress 
and the details will be reported later.

\vskip 1 cm

\noindent {\bf Acknowledgments:}
The library facility provided by Saha Institute of Nuclear Physics, Calcutta, India, is gratefully acknowledged.
Author would also like to thank the referees for their important suggestions.

\newpage

\setlength{\unitlength}{0.240900pt}
\ifx\plotpoint\undefined\newsavebox{\plotpoint}\fi
\sbox{\plotpoint}{\rule[-0.200pt]{0.400pt}{0.400pt}}%
\begin{picture}(1500,1440)(0,0)
\font\gnuplot=cmr10 at 10pt
\gnuplot
\sbox{\plotpoint}{\rule[-0.200pt]{0.400pt}{0.400pt}}%
\put(120.0,123.0){\rule[-0.200pt]{4.818pt}{0.400pt}}
\put(100,123){\makebox(0,0)[r]{0}}
\put(1419.0,123.0){\rule[-0.200pt]{4.818pt}{0.400pt}}
\put(120.0,378.0){\rule[-0.200pt]{4.818pt}{0.400pt}}
\put(100,378){\makebox(0,0)[r]{0.2}}
\put(1419.0,378.0){\rule[-0.200pt]{4.818pt}{0.400pt}}
\put(120.0,634.0){\rule[-0.200pt]{4.818pt}{0.400pt}}
\put(100,634){\makebox(0,0)[r]{0.4}}
\put(1419.0,634.0){\rule[-0.200pt]{4.818pt}{0.400pt}}
\put(120.0,889.0){\rule[-0.200pt]{4.818pt}{0.400pt}}
\put(100,889){\makebox(0,0)[r]{0.6}}
\put(1419.0,889.0){\rule[-0.200pt]{4.818pt}{0.400pt}}
\put(120.0,1145.0){\rule[-0.200pt]{4.818pt}{0.400pt}}
\put(100,1145){\makebox(0,0)[r]{0.8}}
\put(1419.0,1145.0){\rule[-0.200pt]{4.818pt}{0.400pt}}
\put(120.0,1400.0){\rule[-0.200pt]{4.818pt}{0.400pt}}
\put(100,1400){\makebox(0,0)[r]{1}}
\put(1419.0,1400.0){\rule[-0.200pt]{4.818pt}{0.400pt}}
\put(120.0,123.0){\rule[-0.200pt]{0.400pt}{4.818pt}}
\put(120,82){\makebox(0,0){0}}
\put(120.0,1380.0){\rule[-0.200pt]{0.400pt}{4.818pt}}
\put(450.0,123.0){\rule[-0.200pt]{0.400pt}{4.818pt}}
\put(450,82){\makebox(0,0){0.5}}
\put(450.0,1380.0){\rule[-0.200pt]{0.400pt}{4.818pt}}
\put(780.0,123.0){\rule[-0.200pt]{0.400pt}{4.818pt}}
\put(780,82){\makebox(0,0){1}}
\put(780.0,1380.0){\rule[-0.200pt]{0.400pt}{4.818pt}}
\put(1109.0,123.0){\rule[-0.200pt]{0.400pt}{4.818pt}}
\put(1109,82){\makebox(0,0){1.5}}
\put(1109.0,1380.0){\rule[-0.200pt]{0.400pt}{4.818pt}}
\put(1439.0,123.0){\rule[-0.200pt]{0.400pt}{4.818pt}}
\put(1439,82){\makebox(0,0){2}}
\put(1439.0,1380.0){\rule[-0.200pt]{0.400pt}{4.818pt}}
\put(120.0,123.0){\rule[-0.200pt]{317.747pt}{0.400pt}}
\put(1439.0,123.0){\rule[-0.200pt]{0.400pt}{307.629pt}}
\put(120.0,1400.0){\rule[-0.200pt]{317.747pt}{0.400pt}}
\put(779,21){\makebox(0,0){$T$}}
\put(1373,1336){\makebox(0,0){\Large {\bf (a)}}}
\put(1175,1145){\makebox(0,0)[l]{$D=0.2$}}
\put(1175,1017){\makebox(0,0)[l]{$h_{0x} = 0.3$}}
\put(1175,889){\makebox(0,0)[l]{$h_{0z} = 1.0$}}
\put(944,634){\makebox(0,0)[l]{$Q_{y}$}}
\put(648,1145){\makebox(0,0)[l]{$Q_{z}$}}
\put(516,378){\makebox(0,0)[l]{$Q_{x}$}}
\put(120.0,123.0){\rule[-0.200pt]{0.400pt}{307.629pt}}
\put(1307,131){\raisebox{-.8pt}{\makebox(0,0){$\Diamond$}}}
\put(1294,132){\raisebox{-.8pt}{\makebox(0,0){$\Diamond$}}}
\put(1281,132){\raisebox{-.8pt}{\makebox(0,0){$\Diamond$}}}
\put(1268,132){\raisebox{-.8pt}{\makebox(0,0){$\Diamond$}}}
\put(1254,133){\raisebox{-.8pt}{\makebox(0,0){$\Diamond$}}}
\put(1241,133){\raisebox{-.8pt}{\makebox(0,0){$\Diamond$}}}
\put(1228,133){\raisebox{-.8pt}{\makebox(0,0){$\Diamond$}}}
\put(1215,134){\raisebox{-.8pt}{\makebox(0,0){$\Diamond$}}}
\put(1202,134){\raisebox{-.8pt}{\makebox(0,0){$\Diamond$}}}
\put(1188,134){\raisebox{-.8pt}{\makebox(0,0){$\Diamond$}}}
\put(1175,135){\raisebox{-.8pt}{\makebox(0,0){$\Diamond$}}}
\put(1162,136){\raisebox{-.8pt}{\makebox(0,0){$\Diamond$}}}
\put(1149,136){\raisebox{-.8pt}{\makebox(0,0){$\Diamond$}}}
\put(1136,137){\raisebox{-.8pt}{\makebox(0,0){$\Diamond$}}}
\put(1122,137){\raisebox{-.8pt}{\makebox(0,0){$\Diamond$}}}
\put(1109,138){\raisebox{-.8pt}{\makebox(0,0){$\Diamond$}}}
\put(1096,139){\raisebox{-.8pt}{\makebox(0,0){$\Diamond$}}}
\put(1083,140){\raisebox{-.8pt}{\makebox(0,0){$\Diamond$}}}
\put(1070,140){\raisebox{-.8pt}{\makebox(0,0){$\Diamond$}}}
\put(1056,142){\raisebox{-.8pt}{\makebox(0,0){$\Diamond$}}}
\put(1043,143){\raisebox{-.8pt}{\makebox(0,0){$\Diamond$}}}
\put(1030,146){\raisebox{-.8pt}{\makebox(0,0){$\Diamond$}}}
\put(1017,147){\raisebox{-.8pt}{\makebox(0,0){$\Diamond$}}}
\put(1004,152){\raisebox{-.8pt}{\makebox(0,0){$\Diamond$}}}
\put(991,155){\raisebox{-.8pt}{\makebox(0,0){$\Diamond$}}}
\put(977,161){\raisebox{-.8pt}{\makebox(0,0){$\Diamond$}}}
\put(964,170){\raisebox{-.8pt}{\makebox(0,0){$\Diamond$}}}
\put(951,177){\raisebox{-.8pt}{\makebox(0,0){$\Diamond$}}}
\put(938,179){\raisebox{-.8pt}{\makebox(0,0){$\Diamond$}}}
\put(925,189){\raisebox{-.8pt}{\makebox(0,0){$\Diamond$}}}
\put(911,177){\raisebox{-.8pt}{\makebox(0,0){$\Diamond$}}}
\put(898,180){\raisebox{-.8pt}{\makebox(0,0){$\Diamond$}}}
\put(885,181){\raisebox{-.8pt}{\makebox(0,0){$\Diamond$}}}
\put(872,182){\raisebox{-.8pt}{\makebox(0,0){$\Diamond$}}}
\put(859,172){\raisebox{-.8pt}{\makebox(0,0){$\Diamond$}}}
\put(845,177){\raisebox{-.8pt}{\makebox(0,0){$\Diamond$}}}
\put(832,182){\raisebox{-.8pt}{\makebox(0,0){$\Diamond$}}}
\put(819,180){\raisebox{-.8pt}{\makebox(0,0){$\Diamond$}}}
\put(806,184){\raisebox{-.8pt}{\makebox(0,0){$\Diamond$}}}
\put(793,187){\raisebox{-.8pt}{\makebox(0,0){$\Diamond$}}}
\put(780,193){\raisebox{-.8pt}{\makebox(0,0){$\Diamond$}}}
\put(766,213){\raisebox{-.8pt}{\makebox(0,0){$\Diamond$}}}
\put(753,230){\raisebox{-.8pt}{\makebox(0,0){$\Diamond$}}}
\put(740,554){\raisebox{-.8pt}{\makebox(0,0){$\Diamond$}}}
\put(727,583){\raisebox{-.8pt}{\makebox(0,0){$\Diamond$}}}
\put(714,591){\raisebox{-.8pt}{\makebox(0,0){$\Diamond$}}}
\put(700,601){\raisebox{-.8pt}{\makebox(0,0){$\Diamond$}}}
\put(687,541){\raisebox{-.8pt}{\makebox(0,0){$\Diamond$}}}
\put(674,441){\raisebox{-.8pt}{\makebox(0,0){$\Diamond$}}}
\put(661,394){\raisebox{-.8pt}{\makebox(0,0){$\Diamond$}}}
\put(648,372){\raisebox{-.8pt}{\makebox(0,0){$\Diamond$}}}
\put(634,338){\raisebox{-.8pt}{\makebox(0,0){$\Diamond$}}}
\put(621,324){\raisebox{-.8pt}{\makebox(0,0){$\Diamond$}}}
\put(608,303){\raisebox{-.8pt}{\makebox(0,0){$\Diamond$}}}
\put(595,284){\raisebox{-.8pt}{\makebox(0,0){$\Diamond$}}}
\put(582,273){\raisebox{-.8pt}{\makebox(0,0){$\Diamond$}}}
\put(568,263){\raisebox{-.8pt}{\makebox(0,0){$\Diamond$}}}
\put(555,247){\raisebox{-.8pt}{\makebox(0,0){$\Diamond$}}}
\put(542,243){\raisebox{-.8pt}{\makebox(0,0){$\Diamond$}}}
\put(529,236){\raisebox{-.8pt}{\makebox(0,0){$\Diamond$}}}
\put(516,224){\raisebox{-.8pt}{\makebox(0,0){$\Diamond$}}}
\put(503,217){\raisebox{-.8pt}{\makebox(0,0){$\Diamond$}}}
\put(489,212){\raisebox{-.8pt}{\makebox(0,0){$\Diamond$}}}
\put(476,205){\raisebox{-.8pt}{\makebox(0,0){$\Diamond$}}}
\put(463,202){\raisebox{-.8pt}{\makebox(0,0){$\Diamond$}}}
\put(450,196){\raisebox{-.8pt}{\makebox(0,0){$\Diamond$}}}
\put(437,193){\raisebox{-.8pt}{\makebox(0,0){$\Diamond$}}}
\put(423,188){\raisebox{-.8pt}{\makebox(0,0){$\Diamond$}}}
\put(410,183){\raisebox{-.8pt}{\makebox(0,0){$\Diamond$}}}
\put(397,183){\raisebox{-.8pt}{\makebox(0,0){$\Diamond$}}}
\put(384,176){\raisebox{-.8pt}{\makebox(0,0){$\Diamond$}}}
\put(371,168){\raisebox{-.8pt}{\makebox(0,0){$\Diamond$}}}
\put(357,172){\raisebox{-.8pt}{\makebox(0,0){$\Diamond$}}}
\put(344,168){\raisebox{-.8pt}{\makebox(0,0){$\Diamond$}}}
\put(331,166){\raisebox{-.8pt}{\makebox(0,0){$\Diamond$}}}
\put(318,161){\raisebox{-.8pt}{\makebox(0,0){$\Diamond$}}}
\put(305,157){\raisebox{-.8pt}{\makebox(0,0){$\Diamond$}}}
\put(291,150){\raisebox{-.8pt}{\makebox(0,0){$\Diamond$}}}
\put(278,155){\raisebox{-.8pt}{\makebox(0,0){$\Diamond$}}}
\put(265,148){\raisebox{-.8pt}{\makebox(0,0){$\Diamond$}}}
\put(252,149){\raisebox{-.8pt}{\makebox(0,0){$\Diamond$}}}
\put(239,143){\raisebox{-.8pt}{\makebox(0,0){$\Diamond$}}}
\put(226,146){\raisebox{-.8pt}{\makebox(0,0){$\Diamond$}}}
\put(1307,132){\circle{18}}
\put(1294,132){\circle{18}}
\put(1281,133){\circle{18}}
\put(1268,133){\circle{18}}
\put(1254,133){\circle{18}}
\put(1241,134){\circle{18}}
\put(1228,134){\circle{18}}
\put(1215,135){\circle{18}}
\put(1202,135){\circle{18}}
\put(1188,135){\circle{18}}
\put(1175,136){\circle{18}}
\put(1162,136){\circle{18}}
\put(1149,137){\circle{18}}
\put(1136,138){\circle{18}}
\put(1122,138){\circle{18}}
\put(1109,139){\circle{18}}
\put(1096,142){\circle{18}}
\put(1083,142){\circle{18}}
\put(1070,144){\circle{18}}
\put(1056,145){\circle{18}}
\put(1043,148){\circle{18}}
\put(1030,151){\circle{18}}
\put(1017,155){\circle{18}}
\put(1004,158){\circle{18}}
\put(991,164){\circle{18}}
\put(977,177){\circle{18}}
\put(964,186){\circle{18}}
\put(951,243){\circle{18}}
\put(938,358){\circle{18}}
\put(925,463){\circle{18}}
\put(911,539){\circle{18}}
\put(898,586){\circle{18}}
\put(885,629){\circle{18}}
\put(872,665){\circle{18}}
\put(859,697){\circle{18}}
\put(845,723){\circle{18}}
\put(832,748){\circle{18}}
\put(819,772){\circle{18}}
\put(806,791){\circle{18}}
\put(793,811){\circle{18}}
\put(780,825){\circle{18}}
\put(766,838){\circle{18}}
\put(753,840){\circle{18}}
\put(740,320){\circle{18}}
\put(727,231){\circle{18}}
\put(714,221){\circle{18}}
\put(700,202){\circle{18}}
\put(687,169){\circle{18}}
\put(674,164){\circle{18}}
\put(661,158){\circle{18}}
\put(648,155){\circle{18}}
\put(634,151){\circle{18}}
\put(621,149){\circle{18}}
\put(608,149){\circle{18}}
\put(595,148){\circle{18}}
\put(582,148){\circle{18}}
\put(568,147){\circle{18}}
\put(555,145){\circle{18}}
\put(542,147){\circle{18}}
\put(529,142){\circle{18}}
\put(516,142){\circle{18}}
\put(503,142){\circle{18}}
\put(489,142){\circle{18}}
\put(476,142){\circle{18}}
\put(463,140){\circle{18}}
\put(450,138){\circle{18}}
\put(437,140){\circle{18}}
\put(423,140){\circle{18}}
\put(410,139){\circle{18}}
\put(397,138){\circle{18}}
\put(384,137){\circle{18}}
\put(371,137){\circle{18}}
\put(357,137){\circle{18}}
\put(344,135){\circle{18}}
\put(331,136){\circle{18}}
\put(318,135){\circle{18}}
\put(305,134){\circle{18}}
\put(291,134){\circle{18}}
\put(278,134){\circle{18}}
\put(265,133){\circle{18}}
\put(252,133){\circle{18}}
\put(239,130){\circle{18}}
\put(226,131){\circle{18}}
\sbox{\plotpoint}{\rule[-0.500pt]{1.000pt}{1.000pt}}%
\put(1307,130){\circle*{24}}
\put(1294,130){\circle*{24}}
\put(1281,131){\circle*{24}}
\put(1268,131){\circle*{24}}
\put(1254,131){\circle*{24}}
\put(1241,131){\circle*{24}}
\put(1228,131){\circle*{24}}
\put(1215,131){\circle*{24}}
\put(1202,132){\circle*{24}}
\put(1188,132){\circle*{24}}
\put(1175,132){\circle*{24}}
\put(1162,132){\circle*{24}}
\put(1149,133){\circle*{24}}
\put(1136,132){\circle*{24}}
\put(1122,133){\circle*{24}}
\put(1109,133){\circle*{24}}
\put(1096,134){\circle*{24}}
\put(1083,134){\circle*{24}}
\put(1070,135){\circle*{24}}
\put(1056,136){\circle*{24}}
\put(1043,136){\circle*{24}}
\put(1030,137){\circle*{24}}
\put(1017,138){\circle*{24}}
\put(1004,139){\circle*{24}}
\put(991,140){\circle*{24}}
\put(977,142){\circle*{24}}
\put(964,145){\circle*{24}}
\put(951,146){\circle*{24}}
\put(938,149){\circle*{24}}
\put(925,151){\circle*{24}}
\put(911,151){\circle*{24}}
\put(898,155){\circle*{24}}
\put(885,158){\circle*{24}}
\put(872,163){\circle*{24}}
\put(859,163){\circle*{24}}
\put(845,173){\circle*{24}}
\put(832,177){\circle*{24}}
\put(819,185){\circle*{24}}
\put(806,195){\circle*{24}}
\put(793,197){\circle*{24}}
\put(780,225){\circle*{24}}
\put(766,238){\circle*{24}}
\put(753,266){\circle*{24}}
\put(740,694){\circle*{24}}
\put(727,734){\circle*{24}}
\put(714,763){\circle*{24}}
\put(700,829){\circle*{24}}
\put(687,931){\circle*{24}}
\put(674,1009){\circle*{24}}
\put(661,1044){\circle*{24}}
\put(648,1066){\circle*{24}}
\put(634,1089){\circle*{24}}
\put(621,1105){\circle*{24}}
\put(608,1121){\circle*{24}}
\put(595,1136){\circle*{24}}
\put(582,1148){\circle*{24}}
\put(568,1159){\circle*{24}}
\put(555,1171){\circle*{24}}
\put(542,1181){\circle*{24}}
\put(529,1190){\circle*{24}}
\put(516,1200){\circle*{24}}
\put(503,1209){\circle*{24}}
\put(489,1217){\circle*{24}}
\put(476,1226){\circle*{24}}
\put(463,1234){\circle*{24}}
\put(450,1242){\circle*{24}}
\put(437,1249){\circle*{24}}
\put(423,1257){\circle*{24}}
\put(410,1264){\circle*{24}}
\put(397,1271){\circle*{24}}
\put(384,1278){\circle*{24}}
\put(371,1285){\circle*{24}}
\put(357,1292){\circle*{24}}
\put(344,1298){\circle*{24}}
\put(331,1305){\circle*{24}}
\put(318,1311){\circle*{24}}
\put(305,1318){\circle*{24}}
\put(291,1324){\circle*{24}}
\put(278,1330){\circle*{24}}
\put(265,1336){\circle*{24}}
\put(252,1342){\circle*{24}}
\put(239,1348){\circle*{24}}
\put(226,1354){\circle*{24}}
\sbox{\plotpoint}{\rule[-0.200pt]{0.400pt}{0.400pt}}%
\put(1307,131){\usebox{\plotpoint}}
\put(1294,130.67){\rule{3.132pt}{0.400pt}}
\multiput(1300.50,130.17)(-6.500,1.000){2}{\rule{1.566pt}{0.400pt}}
\put(1254,131.67){\rule{3.373pt}{0.400pt}}
\multiput(1261.00,131.17)(-7.000,1.000){2}{\rule{1.686pt}{0.400pt}}
\put(1268.0,132.0){\rule[-0.200pt]{6.263pt}{0.400pt}}
\put(1215,132.67){\rule{3.132pt}{0.400pt}}
\multiput(1221.50,132.17)(-6.500,1.000){2}{\rule{1.566pt}{0.400pt}}
\put(1228.0,133.0){\rule[-0.200pt]{6.263pt}{0.400pt}}
\put(1175,133.67){\rule{3.132pt}{0.400pt}}
\multiput(1181.50,133.17)(-6.500,1.000){2}{\rule{1.566pt}{0.400pt}}
\put(1162,134.67){\rule{3.132pt}{0.400pt}}
\multiput(1168.50,134.17)(-6.500,1.000){2}{\rule{1.566pt}{0.400pt}}
\put(1188.0,134.0){\rule[-0.200pt]{6.504pt}{0.400pt}}
\put(1136,135.67){\rule{3.132pt}{0.400pt}}
\multiput(1142.50,135.17)(-6.500,1.000){2}{\rule{1.566pt}{0.400pt}}
\put(1149.0,136.0){\rule[-0.200pt]{3.132pt}{0.400pt}}
\put(1109,136.67){\rule{3.132pt}{0.400pt}}
\multiput(1115.50,136.17)(-6.500,1.000){2}{\rule{1.566pt}{0.400pt}}
\put(1096,137.67){\rule{3.132pt}{0.400pt}}
\multiput(1102.50,137.17)(-6.500,1.000){2}{\rule{1.566pt}{0.400pt}}
\put(1083,138.67){\rule{3.132pt}{0.400pt}}
\multiput(1089.50,138.17)(-6.500,1.000){2}{\rule{1.566pt}{0.400pt}}
\put(1122.0,137.0){\rule[-0.200pt]{3.373pt}{0.400pt}}
\put(1056,140.17){\rule{2.900pt}{0.400pt}}
\multiput(1063.98,139.17)(-7.981,2.000){2}{\rule{1.450pt}{0.400pt}}
\put(1043,141.67){\rule{3.132pt}{0.400pt}}
\multiput(1049.50,141.17)(-6.500,1.000){2}{\rule{1.566pt}{0.400pt}}
\multiput(1035.39,143.61)(-2.695,0.447){3}{\rule{1.833pt}{0.108pt}}
\multiput(1039.19,142.17)(-9.195,3.000){2}{\rule{0.917pt}{0.400pt}}
\put(1017,145.67){\rule{3.132pt}{0.400pt}}
\multiput(1023.50,145.17)(-6.500,1.000){2}{\rule{1.566pt}{0.400pt}}
\multiput(1012.27,147.59)(-1.378,0.477){7}{\rule{1.140pt}{0.115pt}}
\multiput(1014.63,146.17)(-10.634,5.000){2}{\rule{0.570pt}{0.400pt}}
\multiput(996.39,152.61)(-2.695,0.447){3}{\rule{1.833pt}{0.108pt}}
\multiput(1000.19,151.17)(-9.195,3.000){2}{\rule{0.917pt}{0.400pt}}
\multiput(986.71,155.59)(-1.214,0.482){9}{\rule{1.033pt}{0.116pt}}
\multiput(988.86,154.17)(-11.855,6.000){2}{\rule{0.517pt}{0.400pt}}
\multiput(974.19,161.59)(-0.728,0.489){15}{\rule{0.678pt}{0.118pt}}
\multiput(975.59,160.17)(-11.593,9.000){2}{\rule{0.339pt}{0.400pt}}
\multiput(960.50,170.59)(-0.950,0.485){11}{\rule{0.843pt}{0.117pt}}
\multiput(962.25,169.17)(-11.251,7.000){2}{\rule{0.421pt}{0.400pt}}
\put(938,177.17){\rule{2.700pt}{0.400pt}}
\multiput(945.40,176.17)(-7.396,2.000){2}{\rule{1.350pt}{0.400pt}}
\multiput(935.43,179.58)(-0.652,0.491){17}{\rule{0.620pt}{0.118pt}}
\multiput(936.71,178.17)(-11.713,10.000){2}{\rule{0.310pt}{0.400pt}}
\multiput(922.65,187.92)(-0.582,-0.492){21}{\rule{0.567pt}{0.119pt}}
\multiput(923.82,188.17)(-12.824,-12.000){2}{\rule{0.283pt}{0.400pt}}
\multiput(903.39,177.61)(-2.695,0.447){3}{\rule{1.833pt}{0.108pt}}
\multiput(907.19,176.17)(-9.195,3.000){2}{\rule{0.917pt}{0.400pt}}
\put(885,179.67){\rule{3.132pt}{0.400pt}}
\multiput(891.50,179.17)(-6.500,1.000){2}{\rule{1.566pt}{0.400pt}}
\put(872,180.67){\rule{3.132pt}{0.400pt}}
\multiput(878.50,180.17)(-6.500,1.000){2}{\rule{1.566pt}{0.400pt}}
\multiput(869.43,180.92)(-0.652,-0.491){17}{\rule{0.620pt}{0.118pt}}
\multiput(870.71,181.17)(-11.713,-10.000){2}{\rule{0.310pt}{0.400pt}}
\multiput(853.94,172.59)(-1.489,0.477){7}{\rule{1.220pt}{0.115pt}}
\multiput(856.47,171.17)(-11.468,5.000){2}{\rule{0.610pt}{0.400pt}}
\multiput(840.27,177.59)(-1.378,0.477){7}{\rule{1.140pt}{0.115pt}}
\multiput(842.63,176.17)(-10.634,5.000){2}{\rule{0.570pt}{0.400pt}}
\put(819,180.17){\rule{2.700pt}{0.400pt}}
\multiput(826.40,181.17)(-7.396,-2.000){2}{\rule{1.350pt}{0.400pt}}
\multiput(813.19,180.60)(-1.797,0.468){5}{\rule{1.400pt}{0.113pt}}
\multiput(816.09,179.17)(-10.094,4.000){2}{\rule{0.700pt}{0.400pt}}
\multiput(798.39,184.61)(-2.695,0.447){3}{\rule{1.833pt}{0.108pt}}
\multiput(802.19,183.17)(-9.195,3.000){2}{\rule{0.917pt}{0.400pt}}
\multiput(788.99,187.59)(-1.123,0.482){9}{\rule{0.967pt}{0.116pt}}
\multiput(790.99,186.17)(-10.994,6.000){2}{\rule{0.483pt}{0.400pt}}
\multiput(778.92,193.00)(-0.494,0.717){25}{\rule{0.119pt}{0.671pt}}
\multiput(779.17,193.00)(-14.000,18.606){2}{\rule{0.400pt}{0.336pt}}
\multiput(764.92,213.00)(-0.493,0.655){23}{\rule{0.119pt}{0.623pt}}
\multiput(765.17,213.00)(-13.000,15.707){2}{\rule{0.400pt}{0.312pt}}
\multiput(751.92,230.00)(-0.493,12.827){23}{\rule{0.119pt}{10.069pt}}
\multiput(752.17,230.00)(-13.000,303.101){2}{\rule{0.400pt}{5.035pt}}
\multiput(738.92,554.00)(-0.493,1.131){23}{\rule{0.119pt}{0.992pt}}
\multiput(739.17,554.00)(-13.000,26.940){2}{\rule{0.400pt}{0.496pt}}
\multiput(723.89,583.59)(-0.824,0.488){13}{\rule{0.750pt}{0.117pt}}
\multiput(725.44,582.17)(-11.443,8.000){2}{\rule{0.375pt}{0.400pt}}
\multiput(711.26,591.58)(-0.704,0.491){17}{\rule{0.660pt}{0.118pt}}
\multiput(712.63,590.17)(-12.630,10.000){2}{\rule{0.330pt}{0.400pt}}
\multiput(698.92,592.92)(-0.493,-2.360){23}{\rule{0.119pt}{1.946pt}}
\multiput(699.17,596.96)(-13.000,-55.961){2}{\rule{0.400pt}{0.973pt}}
\multiput(685.92,527.81)(-0.493,-3.946){23}{\rule{0.119pt}{3.177pt}}
\multiput(686.17,534.41)(-13.000,-93.406){2}{\rule{0.400pt}{1.588pt}}
\multiput(672.92,434.58)(-0.493,-1.845){23}{\rule{0.119pt}{1.546pt}}
\multiput(673.17,437.79)(-13.000,-43.791){2}{\rule{0.400pt}{0.773pt}}
\multiput(659.92,390.77)(-0.493,-0.853){23}{\rule{0.119pt}{0.777pt}}
\multiput(660.17,392.39)(-13.000,-20.387){2}{\rule{0.400pt}{0.388pt}}
\multiput(646.92,367.55)(-0.494,-1.231){25}{\rule{0.119pt}{1.071pt}}
\multiput(647.17,369.78)(-14.000,-31.776){2}{\rule{0.400pt}{0.536pt}}
\multiput(632.92,335.80)(-0.493,-0.536){23}{\rule{0.119pt}{0.531pt}}
\multiput(633.17,336.90)(-13.000,-12.898){2}{\rule{0.400pt}{0.265pt}}
\multiput(619.92,320.90)(-0.493,-0.814){23}{\rule{0.119pt}{0.746pt}}
\multiput(620.17,322.45)(-13.000,-19.451){2}{\rule{0.400pt}{0.373pt}}
\multiput(606.92,300.16)(-0.493,-0.734){23}{\rule{0.119pt}{0.685pt}}
\multiput(607.17,301.58)(-13.000,-17.579){2}{\rule{0.400pt}{0.342pt}}
\multiput(592.62,282.92)(-0.590,-0.492){19}{\rule{0.573pt}{0.118pt}}
\multiput(593.81,283.17)(-11.811,-11.000){2}{\rule{0.286pt}{0.400pt}}
\multiput(579.26,271.92)(-0.704,-0.491){17}{\rule{0.660pt}{0.118pt}}
\multiput(580.63,272.17)(-12.630,-10.000){2}{\rule{0.330pt}{0.400pt}}
\multiput(566.92,260.54)(-0.493,-0.616){23}{\rule{0.119pt}{0.592pt}}
\multiput(567.17,261.77)(-13.000,-14.771){2}{\rule{0.400pt}{0.296pt}}
\multiput(549.19,245.94)(-1.797,-0.468){5}{\rule{1.400pt}{0.113pt}}
\multiput(552.09,246.17)(-10.094,-4.000){2}{\rule{0.700pt}{0.400pt}}
\multiput(538.50,241.93)(-0.950,-0.485){11}{\rule{0.843pt}{0.117pt}}
\multiput(540.25,242.17)(-11.251,-7.000){2}{\rule{0.421pt}{0.400pt}}
\multiput(526.79,234.92)(-0.539,-0.492){21}{\rule{0.533pt}{0.119pt}}
\multiput(527.89,235.17)(-11.893,-12.000){2}{\rule{0.267pt}{0.400pt}}
\multiput(512.50,222.93)(-0.950,-0.485){11}{\rule{0.843pt}{0.117pt}}
\multiput(514.25,223.17)(-11.251,-7.000){2}{\rule{0.421pt}{0.400pt}}
\multiput(497.94,215.93)(-1.489,-0.477){7}{\rule{1.220pt}{0.115pt}}
\multiput(500.47,216.17)(-11.468,-5.000){2}{\rule{0.610pt}{0.400pt}}
\multiput(485.50,210.93)(-0.950,-0.485){11}{\rule{0.843pt}{0.117pt}}
\multiput(487.25,211.17)(-11.251,-7.000){2}{\rule{0.421pt}{0.400pt}}
\multiput(468.39,203.95)(-2.695,-0.447){3}{\rule{1.833pt}{0.108pt}}
\multiput(472.19,204.17)(-9.195,-3.000){2}{\rule{0.917pt}{0.400pt}}
\multiput(458.99,200.93)(-1.123,-0.482){9}{\rule{0.967pt}{0.116pt}}
\multiput(460.99,201.17)(-10.994,-6.000){2}{\rule{0.483pt}{0.400pt}}
\multiput(442.39,194.95)(-2.695,-0.447){3}{\rule{1.833pt}{0.108pt}}
\multiput(446.19,195.17)(-9.195,-3.000){2}{\rule{0.917pt}{0.400pt}}
\multiput(431.94,191.93)(-1.489,-0.477){7}{\rule{1.220pt}{0.115pt}}
\multiput(434.47,192.17)(-11.468,-5.000){2}{\rule{0.610pt}{0.400pt}}
\multiput(418.27,186.93)(-1.378,-0.477){7}{\rule{1.140pt}{0.115pt}}
\multiput(420.63,187.17)(-10.634,-5.000){2}{\rule{0.570pt}{0.400pt}}
\put(1070.0,140.0){\rule[-0.200pt]{3.132pt}{0.400pt}}
\multiput(393.50,181.93)(-0.950,-0.485){11}{\rule{0.843pt}{0.117pt}}
\multiput(395.25,182.17)(-11.251,-7.000){2}{\rule{0.421pt}{0.400pt}}
\multiput(380.89,174.93)(-0.824,-0.488){13}{\rule{0.750pt}{0.117pt}}
\multiput(382.44,175.17)(-11.443,-8.000){2}{\rule{0.375pt}{0.400pt}}
\multiput(364.77,168.60)(-1.943,0.468){5}{\rule{1.500pt}{0.113pt}}
\multiput(367.89,167.17)(-10.887,4.000){2}{\rule{0.750pt}{0.400pt}}
\multiput(351.19,170.94)(-1.797,-0.468){5}{\rule{1.400pt}{0.113pt}}
\multiput(354.09,171.17)(-10.094,-4.000){2}{\rule{0.700pt}{0.400pt}}
\put(331,166.17){\rule{2.700pt}{0.400pt}}
\multiput(338.40,167.17)(-7.396,-2.000){2}{\rule{1.350pt}{0.400pt}}
\multiput(326.27,164.93)(-1.378,-0.477){7}{\rule{1.140pt}{0.115pt}}
\multiput(328.63,165.17)(-10.634,-5.000){2}{\rule{0.570pt}{0.400pt}}
\multiput(312.19,159.94)(-1.797,-0.468){5}{\rule{1.400pt}{0.113pt}}
\multiput(315.09,160.17)(-10.094,-4.000){2}{\rule{0.700pt}{0.400pt}}
\multiput(301.26,155.93)(-1.026,-0.485){11}{\rule{0.900pt}{0.117pt}}
\multiput(303.13,156.17)(-12.132,-7.000){2}{\rule{0.450pt}{0.400pt}}
\multiput(286.27,150.59)(-1.378,0.477){7}{\rule{1.140pt}{0.115pt}}
\multiput(288.63,149.17)(-10.634,5.000){2}{\rule{0.570pt}{0.400pt}}
\multiput(274.50,153.93)(-0.950,-0.485){11}{\rule{0.843pt}{0.117pt}}
\multiput(276.25,154.17)(-11.251,-7.000){2}{\rule{0.421pt}{0.400pt}}
\put(252,147.67){\rule{3.132pt}{0.400pt}}
\multiput(258.50,147.17)(-6.500,1.000){2}{\rule{1.566pt}{0.400pt}}
\multiput(247.99,147.93)(-1.123,-0.482){9}{\rule{0.967pt}{0.116pt}}
\multiput(249.99,148.17)(-10.994,-6.000){2}{\rule{0.483pt}{0.400pt}}
\multiput(231.39,143.61)(-2.695,0.447){3}{\rule{1.833pt}{0.108pt}}
\multiput(235.19,142.17)(-9.195,3.000){2}{\rule{0.917pt}{0.400pt}}
\put(397.0,183.0){\rule[-0.200pt]{3.132pt}{0.400pt}}
\put(1307,132){\usebox{\plotpoint}}
\put(1281,131.67){\rule{3.132pt}{0.400pt}}
\multiput(1287.50,131.17)(-6.500,1.000){2}{\rule{1.566pt}{0.400pt}}
\put(1294.0,132.0){\rule[-0.200pt]{3.132pt}{0.400pt}}
\put(1241,132.67){\rule{3.132pt}{0.400pt}}
\multiput(1247.50,132.17)(-6.500,1.000){2}{\rule{1.566pt}{0.400pt}}
\put(1254.0,133.0){\rule[-0.200pt]{6.504pt}{0.400pt}}
\put(1215,133.67){\rule{3.132pt}{0.400pt}}
\multiput(1221.50,133.17)(-6.500,1.000){2}{\rule{1.566pt}{0.400pt}}
\put(1228.0,134.0){\rule[-0.200pt]{3.132pt}{0.400pt}}
\put(1175,134.67){\rule{3.132pt}{0.400pt}}
\multiput(1181.50,134.17)(-6.500,1.000){2}{\rule{1.566pt}{0.400pt}}
\put(1188.0,135.0){\rule[-0.200pt]{6.504pt}{0.400pt}}
\put(1149,135.67){\rule{3.132pt}{0.400pt}}
\multiput(1155.50,135.17)(-6.500,1.000){2}{\rule{1.566pt}{0.400pt}}
\put(1136,136.67){\rule{3.132pt}{0.400pt}}
\multiput(1142.50,136.17)(-6.500,1.000){2}{\rule{1.566pt}{0.400pt}}
\put(1162.0,136.0){\rule[-0.200pt]{3.132pt}{0.400pt}}
\put(1109,137.67){\rule{3.132pt}{0.400pt}}
\multiput(1115.50,137.17)(-6.500,1.000){2}{\rule{1.566pt}{0.400pt}}
\multiput(1101.39,139.61)(-2.695,0.447){3}{\rule{1.833pt}{0.108pt}}
\multiput(1105.19,138.17)(-9.195,3.000){2}{\rule{0.917pt}{0.400pt}}
\put(1122.0,138.0){\rule[-0.200pt]{3.373pt}{0.400pt}}
\put(1070,142.17){\rule{2.700pt}{0.400pt}}
\multiput(1077.40,141.17)(-7.396,2.000){2}{\rule{1.350pt}{0.400pt}}
\put(1056,143.67){\rule{3.373pt}{0.400pt}}
\multiput(1063.00,143.17)(-7.000,1.000){2}{\rule{1.686pt}{0.400pt}}
\multiput(1048.39,145.61)(-2.695,0.447){3}{\rule{1.833pt}{0.108pt}}
\multiput(1052.19,144.17)(-9.195,3.000){2}{\rule{0.917pt}{0.400pt}}
\multiput(1035.39,148.61)(-2.695,0.447){3}{\rule{1.833pt}{0.108pt}}
\multiput(1039.19,147.17)(-9.195,3.000){2}{\rule{0.917pt}{0.400pt}}
\multiput(1024.19,151.60)(-1.797,0.468){5}{\rule{1.400pt}{0.113pt}}
\multiput(1027.09,150.17)(-10.094,4.000){2}{\rule{0.700pt}{0.400pt}}
\multiput(1009.39,155.61)(-2.695,0.447){3}{\rule{1.833pt}{0.108pt}}
\multiput(1013.19,154.17)(-9.195,3.000){2}{\rule{0.917pt}{0.400pt}}
\multiput(999.99,158.59)(-1.123,0.482){9}{\rule{0.967pt}{0.116pt}}
\multiput(1001.99,157.17)(-10.994,6.000){2}{\rule{0.483pt}{0.400pt}}
\multiput(988.80,164.58)(-0.536,0.493){23}{\rule{0.531pt}{0.119pt}}
\multiput(989.90,163.17)(-12.898,13.000){2}{\rule{0.265pt}{0.400pt}}
\multiput(974.19,177.59)(-0.728,0.489){15}{\rule{0.678pt}{0.118pt}}
\multiput(975.59,176.17)(-11.593,9.000){2}{\rule{0.339pt}{0.400pt}}
\multiput(962.92,186.00)(-0.493,2.241){23}{\rule{0.119pt}{1.854pt}}
\multiput(963.17,186.00)(-13.000,53.152){2}{\rule{0.400pt}{0.927pt}}
\multiput(949.92,243.00)(-0.493,4.541){23}{\rule{0.119pt}{3.638pt}}
\multiput(950.17,243.00)(-13.000,107.448){2}{\rule{0.400pt}{1.819pt}}
\multiput(936.92,358.00)(-0.493,4.144){23}{\rule{0.119pt}{3.331pt}}
\multiput(937.17,358.00)(-13.000,98.087){2}{\rule{0.400pt}{1.665pt}}
\multiput(923.92,463.00)(-0.494,2.774){25}{\rule{0.119pt}{2.271pt}}
\multiput(924.17,463.00)(-14.000,71.286){2}{\rule{0.400pt}{1.136pt}}
\multiput(909.92,539.00)(-0.493,1.845){23}{\rule{0.119pt}{1.546pt}}
\multiput(910.17,539.00)(-13.000,43.791){2}{\rule{0.400pt}{0.773pt}}
\multiput(896.92,586.00)(-0.493,1.686){23}{\rule{0.119pt}{1.423pt}}
\multiput(897.17,586.00)(-13.000,40.046){2}{\rule{0.400pt}{0.712pt}}
\multiput(883.92,629.00)(-0.493,1.408){23}{\rule{0.119pt}{1.208pt}}
\multiput(884.17,629.00)(-13.000,33.493){2}{\rule{0.400pt}{0.604pt}}
\multiput(870.92,665.00)(-0.493,1.250){23}{\rule{0.119pt}{1.085pt}}
\multiput(871.17,665.00)(-13.000,29.749){2}{\rule{0.400pt}{0.542pt}}
\multiput(857.92,697.00)(-0.494,0.938){25}{\rule{0.119pt}{0.843pt}}
\multiput(858.17,697.00)(-14.000,24.251){2}{\rule{0.400pt}{0.421pt}}
\multiput(843.92,723.00)(-0.493,0.972){23}{\rule{0.119pt}{0.869pt}}
\multiput(844.17,723.00)(-13.000,23.196){2}{\rule{0.400pt}{0.435pt}}
\multiput(830.92,748.00)(-0.493,0.933){23}{\rule{0.119pt}{0.838pt}}
\multiput(831.17,748.00)(-13.000,22.260){2}{\rule{0.400pt}{0.419pt}}
\multiput(817.92,772.00)(-0.493,0.734){23}{\rule{0.119pt}{0.685pt}}
\multiput(818.17,772.00)(-13.000,17.579){2}{\rule{0.400pt}{0.342pt}}
\multiput(804.92,791.00)(-0.493,0.774){23}{\rule{0.119pt}{0.715pt}}
\multiput(805.17,791.00)(-13.000,18.515){2}{\rule{0.400pt}{0.358pt}}
\multiput(791.92,811.00)(-0.493,0.536){23}{\rule{0.119pt}{0.531pt}}
\multiput(792.17,811.00)(-13.000,12.898){2}{\rule{0.400pt}{0.265pt}}
\multiput(777.80,825.58)(-0.536,0.493){23}{\rule{0.531pt}{0.119pt}}
\multiput(778.90,824.17)(-12.898,13.000){2}{\rule{0.265pt}{0.400pt}}
\put(753,838.17){\rule{2.700pt}{0.400pt}}
\multiput(760.40,837.17)(-7.396,2.000){2}{\rule{1.350pt}{0.400pt}}
\multiput(751.92,773.17)(-0.493,-20.599){23}{\rule{0.119pt}{16.100pt}}
\multiput(752.17,806.58)(-13.000,-486.584){2}{\rule{0.400pt}{8.050pt}}
\multiput(738.92,308.22)(-0.493,-3.510){23}{\rule{0.119pt}{2.838pt}}
\multiput(739.17,314.11)(-13.000,-83.109){2}{\rule{0.400pt}{1.419pt}}
\multiput(724.43,229.92)(-0.652,-0.491){17}{\rule{0.620pt}{0.118pt}}
\multiput(725.71,230.17)(-11.713,-10.000){2}{\rule{0.310pt}{0.400pt}}
\multiput(712.92,218.33)(-0.494,-0.680){25}{\rule{0.119pt}{0.643pt}}
\multiput(713.17,219.67)(-14.000,-17.666){2}{\rule{0.400pt}{0.321pt}}
\multiput(698.92,197.37)(-0.493,-1.290){23}{\rule{0.119pt}{1.115pt}}
\multiput(699.17,199.68)(-13.000,-30.685){2}{\rule{0.400pt}{0.558pt}}
\multiput(682.27,167.93)(-1.378,-0.477){7}{\rule{1.140pt}{0.115pt}}
\multiput(684.63,168.17)(-10.634,-5.000){2}{\rule{0.570pt}{0.400pt}}
\multiput(669.99,162.93)(-1.123,-0.482){9}{\rule{0.967pt}{0.116pt}}
\multiput(671.99,163.17)(-10.994,-6.000){2}{\rule{0.483pt}{0.400pt}}
\multiput(653.39,156.95)(-2.695,-0.447){3}{\rule{1.833pt}{0.108pt}}
\multiput(657.19,157.17)(-9.195,-3.000){2}{\rule{0.917pt}{0.400pt}}
\multiput(641.77,153.94)(-1.943,-0.468){5}{\rule{1.500pt}{0.113pt}}
\multiput(644.89,154.17)(-10.887,-4.000){2}{\rule{0.750pt}{0.400pt}}
\put(621,149.17){\rule{2.700pt}{0.400pt}}
\multiput(628.40,150.17)(-7.396,-2.000){2}{\rule{1.350pt}{0.400pt}}
\put(1083.0,142.0){\rule[-0.200pt]{3.132pt}{0.400pt}}
\put(595,147.67){\rule{3.132pt}{0.400pt}}
\multiput(601.50,148.17)(-6.500,-1.000){2}{\rule{1.566pt}{0.400pt}}
\put(608.0,149.0){\rule[-0.200pt]{3.132pt}{0.400pt}}
\put(568,146.67){\rule{3.373pt}{0.400pt}}
\multiput(575.00,147.17)(-7.000,-1.000){2}{\rule{1.686pt}{0.400pt}}
\put(555,145.17){\rule{2.700pt}{0.400pt}}
\multiput(562.40,146.17)(-7.396,-2.000){2}{\rule{1.350pt}{0.400pt}}
\put(542,145.17){\rule{2.700pt}{0.400pt}}
\multiput(549.40,144.17)(-7.396,2.000){2}{\rule{1.350pt}{0.400pt}}
\multiput(537.27,145.93)(-1.378,-0.477){7}{\rule{1.140pt}{0.115pt}}
\multiput(539.63,146.17)(-10.634,-5.000){2}{\rule{0.570pt}{0.400pt}}
\put(582.0,148.0){\rule[-0.200pt]{3.132pt}{0.400pt}}
\put(463,140.17){\rule{2.700pt}{0.400pt}}
\multiput(470.40,141.17)(-7.396,-2.000){2}{\rule{1.350pt}{0.400pt}}
\put(450,138.17){\rule{2.700pt}{0.400pt}}
\multiput(457.40,139.17)(-7.396,-2.000){2}{\rule{1.350pt}{0.400pt}}
\put(437,138.17){\rule{2.700pt}{0.400pt}}
\multiput(444.40,137.17)(-7.396,2.000){2}{\rule{1.350pt}{0.400pt}}
\put(476.0,142.0){\rule[-0.200pt]{12.768pt}{0.400pt}}
\put(410,138.67){\rule{3.132pt}{0.400pt}}
\multiput(416.50,139.17)(-6.500,-1.000){2}{\rule{1.566pt}{0.400pt}}
\put(397,137.67){\rule{3.132pt}{0.400pt}}
\multiput(403.50,138.17)(-6.500,-1.000){2}{\rule{1.566pt}{0.400pt}}
\put(384,136.67){\rule{3.132pt}{0.400pt}}
\multiput(390.50,137.17)(-6.500,-1.000){2}{\rule{1.566pt}{0.400pt}}
\put(423.0,140.0){\rule[-0.200pt]{3.373pt}{0.400pt}}
\put(344,135.17){\rule{2.700pt}{0.400pt}}
\multiput(351.40,136.17)(-7.396,-2.000){2}{\rule{1.350pt}{0.400pt}}
\put(331,134.67){\rule{3.132pt}{0.400pt}}
\multiput(337.50,134.17)(-6.500,1.000){2}{\rule{1.566pt}{0.400pt}}
\put(318,134.67){\rule{3.132pt}{0.400pt}}
\multiput(324.50,135.17)(-6.500,-1.000){2}{\rule{1.566pt}{0.400pt}}
\put(305,133.67){\rule{3.132pt}{0.400pt}}
\multiput(311.50,134.17)(-6.500,-1.000){2}{\rule{1.566pt}{0.400pt}}
\put(357.0,137.0){\rule[-0.200pt]{6.504pt}{0.400pt}}
\put(265,132.67){\rule{3.132pt}{0.400pt}}
\multiput(271.50,133.17)(-6.500,-1.000){2}{\rule{1.566pt}{0.400pt}}
\put(278.0,134.0){\rule[-0.200pt]{6.504pt}{0.400pt}}
\multiput(244.39,131.95)(-2.695,-0.447){3}{\rule{1.833pt}{0.108pt}}
\multiput(248.19,132.17)(-9.195,-3.000){2}{\rule{0.917pt}{0.400pt}}
\put(226,129.67){\rule{3.132pt}{0.400pt}}
\multiput(232.50,129.17)(-6.500,1.000){2}{\rule{1.566pt}{0.400pt}}
\put(252.0,133.0){\rule[-0.200pt]{3.132pt}{0.400pt}}
\put(1307,130){\usebox{\plotpoint}}
\put(1281,129.67){\rule{3.132pt}{0.400pt}}
\multiput(1287.50,129.17)(-6.500,1.000){2}{\rule{1.566pt}{0.400pt}}
\put(1294.0,130.0){\rule[-0.200pt]{3.132pt}{0.400pt}}
\put(1202,130.67){\rule{3.132pt}{0.400pt}}
\multiput(1208.50,130.17)(-6.500,1.000){2}{\rule{1.566pt}{0.400pt}}
\put(1215.0,131.0){\rule[-0.200pt]{15.899pt}{0.400pt}}
\put(1149,131.67){\rule{3.132pt}{0.400pt}}
\multiput(1155.50,131.17)(-6.500,1.000){2}{\rule{1.566pt}{0.400pt}}
\put(1136,131.67){\rule{3.132pt}{0.400pt}}
\multiput(1142.50,132.17)(-6.500,-1.000){2}{\rule{1.566pt}{0.400pt}}
\put(1122,131.67){\rule{3.373pt}{0.400pt}}
\multiput(1129.00,131.17)(-7.000,1.000){2}{\rule{1.686pt}{0.400pt}}
\put(1162.0,132.0){\rule[-0.200pt]{9.636pt}{0.400pt}}
\put(1096,132.67){\rule{3.132pt}{0.400pt}}
\multiput(1102.50,132.17)(-6.500,1.000){2}{\rule{1.566pt}{0.400pt}}
\put(1109.0,133.0){\rule[-0.200pt]{3.132pt}{0.400pt}}
\put(1070,133.67){\rule{3.132pt}{0.400pt}}
\multiput(1076.50,133.17)(-6.500,1.000){2}{\rule{1.566pt}{0.400pt}}
\put(1056,134.67){\rule{3.373pt}{0.400pt}}
\multiput(1063.00,134.17)(-7.000,1.000){2}{\rule{1.686pt}{0.400pt}}
\put(1083.0,134.0){\rule[-0.200pt]{3.132pt}{0.400pt}}
\put(1030,135.67){\rule{3.132pt}{0.400pt}}
\multiput(1036.50,135.17)(-6.500,1.000){2}{\rule{1.566pt}{0.400pt}}
\put(1017,136.67){\rule{3.132pt}{0.400pt}}
\multiput(1023.50,136.17)(-6.500,1.000){2}{\rule{1.566pt}{0.400pt}}
\put(1004,137.67){\rule{3.132pt}{0.400pt}}
\multiput(1010.50,137.17)(-6.500,1.000){2}{\rule{1.566pt}{0.400pt}}
\put(991,138.67){\rule{3.132pt}{0.400pt}}
\multiput(997.50,138.17)(-6.500,1.000){2}{\rule{1.566pt}{0.400pt}}
\put(977,140.17){\rule{2.900pt}{0.400pt}}
\multiput(984.98,139.17)(-7.981,2.000){2}{\rule{1.450pt}{0.400pt}}
\multiput(969.39,142.61)(-2.695,0.447){3}{\rule{1.833pt}{0.108pt}}
\multiput(973.19,141.17)(-9.195,3.000){2}{\rule{0.917pt}{0.400pt}}
\put(951,144.67){\rule{3.132pt}{0.400pt}}
\multiput(957.50,144.17)(-6.500,1.000){2}{\rule{1.566pt}{0.400pt}}
\multiput(943.39,146.61)(-2.695,0.447){3}{\rule{1.833pt}{0.108pt}}
\multiput(947.19,145.17)(-9.195,3.000){2}{\rule{0.917pt}{0.400pt}}
\put(925,149.17){\rule{2.700pt}{0.400pt}}
\multiput(932.40,148.17)(-7.396,2.000){2}{\rule{1.350pt}{0.400pt}}
\put(1043.0,136.0){\rule[-0.200pt]{3.132pt}{0.400pt}}
\multiput(905.19,151.60)(-1.797,0.468){5}{\rule{1.400pt}{0.113pt}}
\multiput(908.09,150.17)(-10.094,4.000){2}{\rule{0.700pt}{0.400pt}}
\multiput(890.39,155.61)(-2.695,0.447){3}{\rule{1.833pt}{0.108pt}}
\multiput(894.19,154.17)(-9.195,3.000){2}{\rule{0.917pt}{0.400pt}}
\multiput(880.27,158.59)(-1.378,0.477){7}{\rule{1.140pt}{0.115pt}}
\multiput(882.63,157.17)(-10.634,5.000){2}{\rule{0.570pt}{0.400pt}}
\put(911.0,151.0){\rule[-0.200pt]{3.373pt}{0.400pt}}
\multiput(856.26,163.58)(-0.704,0.491){17}{\rule{0.660pt}{0.118pt}}
\multiput(857.63,162.17)(-12.630,10.000){2}{\rule{0.330pt}{0.400pt}}
\multiput(839.19,173.60)(-1.797,0.468){5}{\rule{1.400pt}{0.113pt}}
\multiput(842.09,172.17)(-10.094,4.000){2}{\rule{0.700pt}{0.400pt}}
\multiput(828.89,177.59)(-0.824,0.488){13}{\rule{0.750pt}{0.117pt}}
\multiput(830.44,176.17)(-11.443,8.000){2}{\rule{0.375pt}{0.400pt}}
\multiput(816.43,185.58)(-0.652,0.491){17}{\rule{0.620pt}{0.118pt}}
\multiput(817.71,184.17)(-11.713,10.000){2}{\rule{0.310pt}{0.400pt}}
\put(793,195.17){\rule{2.700pt}{0.400pt}}
\multiput(800.40,194.17)(-7.396,2.000){2}{\rule{1.350pt}{0.400pt}}
\multiput(791.92,197.00)(-0.493,1.091){23}{\rule{0.119pt}{0.962pt}}
\multiput(792.17,197.00)(-13.000,26.004){2}{\rule{0.400pt}{0.481pt}}
\multiput(777.80,225.58)(-0.536,0.493){23}{\rule{0.531pt}{0.119pt}}
\multiput(778.90,224.17)(-12.898,13.000){2}{\rule{0.265pt}{0.400pt}}
\multiput(764.92,238.00)(-0.493,1.091){23}{\rule{0.119pt}{0.962pt}}
\multiput(765.17,238.00)(-13.000,26.004){2}{\rule{0.400pt}{0.481pt}}
\multiput(751.92,266.00)(-0.493,16.951){23}{\rule{0.119pt}{13.269pt}}
\multiput(752.17,266.00)(-13.000,400.459){2}{\rule{0.400pt}{6.635pt}}
\multiput(738.92,694.00)(-0.493,1.567){23}{\rule{0.119pt}{1.331pt}}
\multiput(739.17,694.00)(-13.000,37.238){2}{\rule{0.400pt}{0.665pt}}
\multiput(725.92,734.00)(-0.493,1.131){23}{\rule{0.119pt}{0.992pt}}
\multiput(726.17,734.00)(-13.000,26.940){2}{\rule{0.400pt}{0.496pt}}
\multiput(712.92,763.00)(-0.494,2.407){25}{\rule{0.119pt}{1.986pt}}
\multiput(713.17,763.00)(-14.000,61.879){2}{\rule{0.400pt}{0.993pt}}
\multiput(698.92,829.00)(-0.493,4.025){23}{\rule{0.119pt}{3.238pt}}
\multiput(699.17,829.00)(-13.000,95.278){2}{\rule{0.400pt}{1.619pt}}
\multiput(685.92,931.00)(-0.493,3.074){23}{\rule{0.119pt}{2.500pt}}
\multiput(686.17,931.00)(-13.000,72.811){2}{\rule{0.400pt}{1.250pt}}
\multiput(672.92,1009.00)(-0.493,1.369){23}{\rule{0.119pt}{1.177pt}}
\multiput(673.17,1009.00)(-13.000,32.557){2}{\rule{0.400pt}{0.588pt}}
\multiput(659.92,1044.00)(-0.493,0.853){23}{\rule{0.119pt}{0.777pt}}
\multiput(660.17,1044.00)(-13.000,20.387){2}{\rule{0.400pt}{0.388pt}}
\multiput(646.92,1066.00)(-0.494,0.827){25}{\rule{0.119pt}{0.757pt}}
\multiput(647.17,1066.00)(-14.000,21.429){2}{\rule{0.400pt}{0.379pt}}
\multiput(632.92,1089.00)(-0.493,0.616){23}{\rule{0.119pt}{0.592pt}}
\multiput(633.17,1089.00)(-13.000,14.771){2}{\rule{0.400pt}{0.296pt}}
\multiput(619.92,1105.00)(-0.493,0.616){23}{\rule{0.119pt}{0.592pt}}
\multiput(620.17,1105.00)(-13.000,14.771){2}{\rule{0.400pt}{0.296pt}}
\multiput(606.92,1121.00)(-0.493,0.576){23}{\rule{0.119pt}{0.562pt}}
\multiput(607.17,1121.00)(-13.000,13.834){2}{\rule{0.400pt}{0.281pt}}
\multiput(592.79,1136.58)(-0.539,0.492){21}{\rule{0.533pt}{0.119pt}}
\multiput(593.89,1135.17)(-11.893,12.000){2}{\rule{0.267pt}{0.400pt}}
\multiput(579.47,1148.58)(-0.637,0.492){19}{\rule{0.609pt}{0.118pt}}
\multiput(580.74,1147.17)(-12.736,11.000){2}{\rule{0.305pt}{0.400pt}}
\multiput(565.79,1159.58)(-0.539,0.492){21}{\rule{0.533pt}{0.119pt}}
\multiput(566.89,1158.17)(-11.893,12.000){2}{\rule{0.267pt}{0.400pt}}
\multiput(552.43,1171.58)(-0.652,0.491){17}{\rule{0.620pt}{0.118pt}}
\multiput(553.71,1170.17)(-11.713,10.000){2}{\rule{0.310pt}{0.400pt}}
\multiput(539.19,1181.59)(-0.728,0.489){15}{\rule{0.678pt}{0.118pt}}
\multiput(540.59,1180.17)(-11.593,9.000){2}{\rule{0.339pt}{0.400pt}}
\multiput(526.43,1190.58)(-0.652,0.491){17}{\rule{0.620pt}{0.118pt}}
\multiput(527.71,1189.17)(-11.713,10.000){2}{\rule{0.310pt}{0.400pt}}
\multiput(513.19,1200.59)(-0.728,0.489){15}{\rule{0.678pt}{0.118pt}}
\multiput(514.59,1199.17)(-11.593,9.000){2}{\rule{0.339pt}{0.400pt}}
\multiput(499.68,1209.59)(-0.890,0.488){13}{\rule{0.800pt}{0.117pt}}
\multiput(501.34,1208.17)(-12.340,8.000){2}{\rule{0.400pt}{0.400pt}}
\multiput(486.19,1217.59)(-0.728,0.489){15}{\rule{0.678pt}{0.118pt}}
\multiput(487.59,1216.17)(-11.593,9.000){2}{\rule{0.339pt}{0.400pt}}
\multiput(472.89,1226.59)(-0.824,0.488){13}{\rule{0.750pt}{0.117pt}}
\multiput(474.44,1225.17)(-11.443,8.000){2}{\rule{0.375pt}{0.400pt}}
\multiput(459.89,1234.59)(-0.824,0.488){13}{\rule{0.750pt}{0.117pt}}
\multiput(461.44,1233.17)(-11.443,8.000){2}{\rule{0.375pt}{0.400pt}}
\multiput(446.50,1242.59)(-0.950,0.485){11}{\rule{0.843pt}{0.117pt}}
\multiput(448.25,1241.17)(-11.251,7.000){2}{\rule{0.421pt}{0.400pt}}
\multiput(433.68,1249.59)(-0.890,0.488){13}{\rule{0.800pt}{0.117pt}}
\multiput(435.34,1248.17)(-12.340,8.000){2}{\rule{0.400pt}{0.400pt}}
\multiput(419.50,1257.59)(-0.950,0.485){11}{\rule{0.843pt}{0.117pt}}
\multiput(421.25,1256.17)(-11.251,7.000){2}{\rule{0.421pt}{0.400pt}}
\multiput(406.50,1264.59)(-0.950,0.485){11}{\rule{0.843pt}{0.117pt}}
\multiput(408.25,1263.17)(-11.251,7.000){2}{\rule{0.421pt}{0.400pt}}
\multiput(393.50,1271.59)(-0.950,0.485){11}{\rule{0.843pt}{0.117pt}}
\multiput(395.25,1270.17)(-11.251,7.000){2}{\rule{0.421pt}{0.400pt}}
\multiput(380.50,1278.59)(-0.950,0.485){11}{\rule{0.843pt}{0.117pt}}
\multiput(382.25,1277.17)(-11.251,7.000){2}{\rule{0.421pt}{0.400pt}}
\multiput(367.26,1285.59)(-1.026,0.485){11}{\rule{0.900pt}{0.117pt}}
\multiput(369.13,1284.17)(-12.132,7.000){2}{\rule{0.450pt}{0.400pt}}
\multiput(352.99,1292.59)(-1.123,0.482){9}{\rule{0.967pt}{0.116pt}}
\multiput(354.99,1291.17)(-10.994,6.000){2}{\rule{0.483pt}{0.400pt}}
\multiput(340.50,1298.59)(-0.950,0.485){11}{\rule{0.843pt}{0.117pt}}
\multiput(342.25,1297.17)(-11.251,7.000){2}{\rule{0.421pt}{0.400pt}}
\multiput(326.99,1305.59)(-1.123,0.482){9}{\rule{0.967pt}{0.116pt}}
\multiput(328.99,1304.17)(-10.994,6.000){2}{\rule{0.483pt}{0.400pt}}
\multiput(314.50,1311.59)(-0.950,0.485){11}{\rule{0.843pt}{0.117pt}}
\multiput(316.25,1310.17)(-11.251,7.000){2}{\rule{0.421pt}{0.400pt}}
\multiput(300.71,1318.59)(-1.214,0.482){9}{\rule{1.033pt}{0.116pt}}
\multiput(302.86,1317.17)(-11.855,6.000){2}{\rule{0.517pt}{0.400pt}}
\multiput(286.99,1324.59)(-1.123,0.482){9}{\rule{0.967pt}{0.116pt}}
\multiput(288.99,1323.17)(-10.994,6.000){2}{\rule{0.483pt}{0.400pt}}
\multiput(273.99,1330.59)(-1.123,0.482){9}{\rule{0.967pt}{0.116pt}}
\multiput(275.99,1329.17)(-10.994,6.000){2}{\rule{0.483pt}{0.400pt}}
\multiput(260.99,1336.59)(-1.123,0.482){9}{\rule{0.967pt}{0.116pt}}
\multiput(262.99,1335.17)(-10.994,6.000){2}{\rule{0.483pt}{0.400pt}}
\multiput(247.99,1342.59)(-1.123,0.482){9}{\rule{0.967pt}{0.116pt}}
\multiput(249.99,1341.17)(-10.994,6.000){2}{\rule{0.483pt}{0.400pt}}
\multiput(234.99,1348.59)(-1.123,0.482){9}{\rule{0.967pt}{0.116pt}}
\multiput(236.99,1347.17)(-10.994,6.000){2}{\rule{0.483pt}{0.400pt}}
\put(859.0,163.0){\rule[-0.200pt]{3.132pt}{0.400pt}}
\end{picture}

\begin{center}{\Large {\bf Fig.1(a)}}\end{center}

\noindent {\bf Fig.1(a):} The temperature variations of the components of dynamic order parameters. Different
symbols represent different components. 
$Q_x$ $({\Diamond})$, $Q_y$ ( {\circle{18}}) and $Q_z$ ( {\circle*{24}}). This
diagram is for $D = 0.2$ and for elliptically polarised field where $h_{0x}=0.3$ and $h_{0z}=1.0$. 
The size of the errorbars of $Q_x$, $Q_y$ and $Q_z$ close to the transition points is of the order of 0.02 and 
that at low temperature (e.g., below $T = 0.5$) is around 0.003.  
\newpage

\setlength{\unitlength}{0.240900pt}
\ifx\plotpoint\undefined\newsavebox{\plotpoint}\fi
\sbox{\plotpoint}{\rule[-0.200pt]{0.400pt}{0.400pt}}%
\begin{picture}(1500,1440)(0,0)
\font\gnuplot=cmr10 at 10pt
\gnuplot
\sbox{\plotpoint}{\rule[-0.200pt]{0.400pt}{0.400pt}}%
\put(181.0,123.0){\rule[-0.200pt]{4.818pt}{0.400pt}}
\put(161,123){\makebox(0,0)[r]{-3.2}}
\put(1419.0,123.0){\rule[-0.200pt]{4.818pt}{0.400pt}}
\put(181.0,239.0){\rule[-0.200pt]{4.818pt}{0.400pt}}
\put(161,239){\makebox(0,0)[r]{-3}}
\put(1419.0,239.0){\rule[-0.200pt]{4.818pt}{0.400pt}}
\put(181.0,355.0){\rule[-0.200pt]{4.818pt}{0.400pt}}
\put(161,355){\makebox(0,0)[r]{-2.8}}
\put(1419.0,355.0){\rule[-0.200pt]{4.818pt}{0.400pt}}
\put(181.0,471.0){\rule[-0.200pt]{4.818pt}{0.400pt}}
\put(161,471){\makebox(0,0)[r]{-2.6}}
\put(1419.0,471.0){\rule[-0.200pt]{4.818pt}{0.400pt}}
\put(181.0,587.0){\rule[-0.200pt]{4.818pt}{0.400pt}}
\put(161,587){\makebox(0,0)[r]{-2.4}}
\put(1419.0,587.0){\rule[-0.200pt]{4.818pt}{0.400pt}}
\put(181.0,703.0){\rule[-0.200pt]{4.818pt}{0.400pt}}
\put(161,703){\makebox(0,0)[r]{-2.2}}
\put(1419.0,703.0){\rule[-0.200pt]{4.818pt}{0.400pt}}
\put(181.0,820.0){\rule[-0.200pt]{4.818pt}{0.400pt}}
\put(161,820){\makebox(0,0)[r]{-2}}
\put(1419.0,820.0){\rule[-0.200pt]{4.818pt}{0.400pt}}
\put(181.0,936.0){\rule[-0.200pt]{4.818pt}{0.400pt}}
\put(161,936){\makebox(0,0)[r]{-1.8}}
\put(1419.0,936.0){\rule[-0.200pt]{4.818pt}{0.400pt}}
\put(181.0,1052.0){\rule[-0.200pt]{4.818pt}{0.400pt}}
\put(161,1052){\makebox(0,0)[r]{-1.6}}
\put(1419.0,1052.0){\rule[-0.200pt]{4.818pt}{0.400pt}}
\put(181.0,1168.0){\rule[-0.200pt]{4.818pt}{0.400pt}}
\put(161,1168){\makebox(0,0)[r]{-1.4}}
\put(1419.0,1168.0){\rule[-0.200pt]{4.818pt}{0.400pt}}
\put(181.0,1284.0){\rule[-0.200pt]{4.818pt}{0.400pt}}
\put(161,1284){\makebox(0,0)[r]{-1.2}}
\put(1419.0,1284.0){\rule[-0.200pt]{4.818pt}{0.400pt}}
\put(181.0,1400.0){\rule[-0.200pt]{4.818pt}{0.400pt}}
\put(161,1400){\makebox(0,0)[r]{-1}}
\put(1419.0,1400.0){\rule[-0.200pt]{4.818pt}{0.400pt}}
\put(181.0,123.0){\rule[-0.200pt]{0.400pt}{4.818pt}}
\put(181,82){\makebox(0,0){0}}
\put(181.0,1380.0){\rule[-0.200pt]{0.400pt}{4.818pt}}
\put(496.0,123.0){\rule[-0.200pt]{0.400pt}{4.818pt}}
\put(496,82){\makebox(0,0){0.5}}
\put(496.0,1380.0){\rule[-0.200pt]{0.400pt}{4.818pt}}
\put(810.0,123.0){\rule[-0.200pt]{0.400pt}{4.818pt}}
\put(810,82){\makebox(0,0){1}}
\put(810.0,1380.0){\rule[-0.200pt]{0.400pt}{4.818pt}}
\put(1125.0,123.0){\rule[-0.200pt]{0.400pt}{4.818pt}}
\put(1125,82){\makebox(0,0){1.5}}
\put(1125.0,1380.0){\rule[-0.200pt]{0.400pt}{4.818pt}}
\put(1439.0,123.0){\rule[-0.200pt]{0.400pt}{4.818pt}}
\put(1439,82){\makebox(0,0){2}}
\put(1439.0,1380.0){\rule[-0.200pt]{0.400pt}{4.818pt}}
\put(181.0,123.0){\rule[-0.200pt]{303.052pt}{0.400pt}}
\put(1439.0,123.0){\rule[-0.200pt]{0.400pt}{307.629pt}}
\put(181.0,1400.0){\rule[-0.200pt]{303.052pt}{0.400pt}}
\put(40,761){\makebox(0,0){$E$}}
\put(810,21){\makebox(0,0){$T$}}
\put(1376,1342){\makebox(0,0){\Large {\bf (b)}}}
\put(1125,820){\makebox(0,0)[l]{$D=0.2$}}
\put(1125,703){\makebox(0,0)[l]{$h_{0x} = 0.3$}}
\put(1125,587){\makebox(0,0)[l]{$h_{0z} = 1.0$}}
\put(181.0,123.0){\rule[-0.200pt]{0.400pt}{307.629pt}}
\sbox{\plotpoint}{\rule[-0.600pt]{1.200pt}{1.200pt}}%
\put(1313,1344){\circle*{18}}
\put(1301,1338){\circle*{18}}
\put(1288,1334){\circle*{18}}
\put(1275,1330){\circle*{18}}
\put(1263,1325){\circle*{18}}
\put(1250,1321){\circle*{18}}
\put(1238,1317){\circle*{18}}
\put(1225,1313){\circle*{18}}
\put(1213,1309){\circle*{18}}
\put(1200,1305){\circle*{18}}
\put(1187,1302){\circle*{18}}
\put(1175,1299){\circle*{18}}
\put(1162,1295){\circle*{18}}
\put(1150,1293){\circle*{18}}
\put(1137,1290){\circle*{18}}
\put(1125,1287){\circle*{18}}
\put(1112,1284){\circle*{18}}
\put(1099,1282){\circle*{18}}
\put(1087,1279){\circle*{18}}
\put(1074,1276){\circle*{18}}
\put(1062,1273){\circle*{18}}
\put(1049,1271){\circle*{18}}
\put(1036,1267){\circle*{18}}
\put(1024,1263){\circle*{18}}
\put(1011,1259){\circle*{18}}
\put(999,1253){\circle*{18}}
\put(986,1246){\circle*{18}}
\put(974,1235){\circle*{18}}
\put(961,1213){\circle*{18}}
\put(948,1186){\circle*{18}}
\put(936,1160){\circle*{18}}
\put(923,1136){\circle*{18}}
\put(911,1114){\circle*{18}}
\put(898,1092){\circle*{18}}
\put(885,1072){\circle*{18}}
\put(873,1052){\circle*{18}}
\put(860,1032){\circle*{18}}
\put(848,1013){\circle*{18}}
\put(835,994){\circle*{18}}
\put(823,976){\circle*{18}}
\put(810,958){\circle*{18}}
\put(797,940){\circle*{18}}
\put(785,921){\circle*{18}}
\put(772,854){\circle*{18}}
\put(760,831){\circle*{18}}
\put(747,810){\circle*{18}}
\put(735,786){\circle*{18}}
\put(722,752){\circle*{18}}
\put(709,719){\circle*{18}}
\put(697,696){\circle*{18}}
\put(684,677){\circle*{18}}
\put(672,657){\circle*{18}}
\put(659,640){\circle*{18}}
\put(646,622){\circle*{18}}
\put(634,606){\circle*{18}}
\put(621,589){\circle*{18}}
\put(609,574){\circle*{18}}
\put(596,558){\circle*{18}}
\put(584,543){\circle*{18}}
\put(571,529){\circle*{18}}
\put(558,514){\circle*{18}}
\put(546,500){\circle*{18}}
\put(533,485){\circle*{18}}
\put(521,471){\circle*{18}}
\put(508,457){\circle*{18}}
\put(496,444){\circle*{18}}
\put(483,430){\circle*{18}}
\put(470,416){\circle*{18}}
\put(458,403){\circle*{18}}
\put(445,390){\circle*{18}}
\put(433,377){\circle*{18}}
\put(420,363){\circle*{18}}
\put(407,350){\circle*{18}}
\put(395,337){\circle*{18}}
\put(382,325){\circle*{18}}
\put(370,312){\circle*{18}}
\put(357,299){\circle*{18}}
\put(345,286){\circle*{18}}
\put(332,274){\circle*{18}}
\put(319,261){\circle*{18}}
\put(307,249){\circle*{18}}
\put(294,236){\circle*{18}}
\put(282,224){\circle*{18}}
\sbox{\plotpoint}{\rule[-0.200pt]{0.400pt}{0.400pt}}%
\put(1313,1344){\usebox{\plotpoint}}
\multiput(1309.26,1342.93)(-1.033,-0.482){9}{\rule{0.900pt}{0.116pt}}
\multiput(1311.13,1343.17)(-10.132,-6.000){2}{\rule{0.450pt}{0.400pt}}
\multiput(1295.19,1336.94)(-1.797,-0.468){5}{\rule{1.400pt}{0.113pt}}
\multiput(1298.09,1337.17)(-10.094,-4.000){2}{\rule{0.700pt}{0.400pt}}
\multiput(1282.19,1332.94)(-1.797,-0.468){5}{\rule{1.400pt}{0.113pt}}
\multiput(1285.09,1333.17)(-10.094,-4.000){2}{\rule{0.700pt}{0.400pt}}
\multiput(1270.60,1328.93)(-1.267,-0.477){7}{\rule{1.060pt}{0.115pt}}
\multiput(1272.80,1329.17)(-9.800,-5.000){2}{\rule{0.530pt}{0.400pt}}
\multiput(1257.19,1323.94)(-1.797,-0.468){5}{\rule{1.400pt}{0.113pt}}
\multiput(1260.09,1324.17)(-10.094,-4.000){2}{\rule{0.700pt}{0.400pt}}
\multiput(1244.60,1319.94)(-1.651,-0.468){5}{\rule{1.300pt}{0.113pt}}
\multiput(1247.30,1320.17)(-9.302,-4.000){2}{\rule{0.650pt}{0.400pt}}
\multiput(1232.19,1315.94)(-1.797,-0.468){5}{\rule{1.400pt}{0.113pt}}
\multiput(1235.09,1316.17)(-10.094,-4.000){2}{\rule{0.700pt}{0.400pt}}
\multiput(1219.60,1311.94)(-1.651,-0.468){5}{\rule{1.300pt}{0.113pt}}
\multiput(1222.30,1312.17)(-9.302,-4.000){2}{\rule{0.650pt}{0.400pt}}
\multiput(1207.19,1307.94)(-1.797,-0.468){5}{\rule{1.400pt}{0.113pt}}
\multiput(1210.09,1308.17)(-10.094,-4.000){2}{\rule{0.700pt}{0.400pt}}
\multiput(1192.39,1303.95)(-2.695,-0.447){3}{\rule{1.833pt}{0.108pt}}
\multiput(1196.19,1304.17)(-9.195,-3.000){2}{\rule{0.917pt}{0.400pt}}
\multiput(1179.94,1300.95)(-2.472,-0.447){3}{\rule{1.700pt}{0.108pt}}
\multiput(1183.47,1301.17)(-8.472,-3.000){2}{\rule{0.850pt}{0.400pt}}
\multiput(1169.19,1297.94)(-1.797,-0.468){5}{\rule{1.400pt}{0.113pt}}
\multiput(1172.09,1298.17)(-10.094,-4.000){2}{\rule{0.700pt}{0.400pt}}
\put(1150,1293.17){\rule{2.500pt}{0.400pt}}
\multiput(1156.81,1294.17)(-6.811,-2.000){2}{\rule{1.250pt}{0.400pt}}
\multiput(1142.39,1291.95)(-2.695,-0.447){3}{\rule{1.833pt}{0.108pt}}
\multiput(1146.19,1292.17)(-9.195,-3.000){2}{\rule{0.917pt}{0.400pt}}
\multiput(1129.94,1288.95)(-2.472,-0.447){3}{\rule{1.700pt}{0.108pt}}
\multiput(1133.47,1289.17)(-8.472,-3.000){2}{\rule{0.850pt}{0.400pt}}
\multiput(1117.39,1285.95)(-2.695,-0.447){3}{\rule{1.833pt}{0.108pt}}
\multiput(1121.19,1286.17)(-9.195,-3.000){2}{\rule{0.917pt}{0.400pt}}
\put(1099,1282.17){\rule{2.700pt}{0.400pt}}
\multiput(1106.40,1283.17)(-7.396,-2.000){2}{\rule{1.350pt}{0.400pt}}
\multiput(1091.94,1280.95)(-2.472,-0.447){3}{\rule{1.700pt}{0.108pt}}
\multiput(1095.47,1281.17)(-8.472,-3.000){2}{\rule{0.850pt}{0.400pt}}
\multiput(1079.39,1277.95)(-2.695,-0.447){3}{\rule{1.833pt}{0.108pt}}
\multiput(1083.19,1278.17)(-9.195,-3.000){2}{\rule{0.917pt}{0.400pt}}
\multiput(1066.94,1274.95)(-2.472,-0.447){3}{\rule{1.700pt}{0.108pt}}
\multiput(1070.47,1275.17)(-8.472,-3.000){2}{\rule{0.850pt}{0.400pt}}
\put(1049,1271.17){\rule{2.700pt}{0.400pt}}
\multiput(1056.40,1272.17)(-7.396,-2.000){2}{\rule{1.350pt}{0.400pt}}
\multiput(1043.19,1269.94)(-1.797,-0.468){5}{\rule{1.400pt}{0.113pt}}
\multiput(1046.09,1270.17)(-10.094,-4.000){2}{\rule{0.700pt}{0.400pt}}
\multiput(1030.60,1265.94)(-1.651,-0.468){5}{\rule{1.300pt}{0.113pt}}
\multiput(1033.30,1266.17)(-9.302,-4.000){2}{\rule{0.650pt}{0.400pt}}
\multiput(1018.19,1261.94)(-1.797,-0.468){5}{\rule{1.400pt}{0.113pt}}
\multiput(1021.09,1262.17)(-10.094,-4.000){2}{\rule{0.700pt}{0.400pt}}
\multiput(1007.26,1257.93)(-1.033,-0.482){9}{\rule{0.900pt}{0.116pt}}
\multiput(1009.13,1258.17)(-10.132,-6.000){2}{\rule{0.450pt}{0.400pt}}
\multiput(995.50,1251.93)(-0.950,-0.485){11}{\rule{0.843pt}{0.117pt}}
\multiput(997.25,1252.17)(-11.251,-7.000){2}{\rule{0.421pt}{0.400pt}}
\multiput(983.77,1244.92)(-0.543,-0.492){19}{\rule{0.536pt}{0.118pt}}
\multiput(984.89,1245.17)(-10.887,-11.000){2}{\rule{0.268pt}{0.400pt}}
\multiput(972.92,1231.77)(-0.493,-0.853){23}{\rule{0.119pt}{0.777pt}}
\multiput(973.17,1233.39)(-13.000,-20.387){2}{\rule{0.400pt}{0.388pt}}
\multiput(959.92,1209.14)(-0.493,-1.052){23}{\rule{0.119pt}{0.931pt}}
\multiput(960.17,1211.07)(-13.000,-25.068){2}{\rule{0.400pt}{0.465pt}}
\multiput(946.92,1181.99)(-0.492,-1.099){21}{\rule{0.119pt}{0.967pt}}
\multiput(947.17,1183.99)(-12.000,-23.994){2}{\rule{0.400pt}{0.483pt}}
\multiput(934.92,1156.52)(-0.493,-0.933){23}{\rule{0.119pt}{0.838pt}}
\multiput(935.17,1158.26)(-13.000,-22.260){2}{\rule{0.400pt}{0.419pt}}
\multiput(921.92,1132.54)(-0.492,-0.927){21}{\rule{0.119pt}{0.833pt}}
\multiput(922.17,1134.27)(-12.000,-20.270){2}{\rule{0.400pt}{0.417pt}}
\multiput(909.92,1110.77)(-0.493,-0.853){23}{\rule{0.119pt}{0.777pt}}
\multiput(910.17,1112.39)(-13.000,-20.387){2}{\rule{0.400pt}{0.388pt}}
\multiput(896.92,1089.03)(-0.493,-0.774){23}{\rule{0.119pt}{0.715pt}}
\multiput(897.17,1090.52)(-13.000,-18.515){2}{\rule{0.400pt}{0.358pt}}
\multiput(883.92,1068.82)(-0.492,-0.841){21}{\rule{0.119pt}{0.767pt}}
\multiput(884.17,1070.41)(-12.000,-18.409){2}{\rule{0.400pt}{0.383pt}}
\multiput(871.92,1049.03)(-0.493,-0.774){23}{\rule{0.119pt}{0.715pt}}
\multiput(872.17,1050.52)(-13.000,-18.515){2}{\rule{0.400pt}{0.358pt}}
\multiput(858.92,1028.96)(-0.492,-0.798){21}{\rule{0.119pt}{0.733pt}}
\multiput(859.17,1030.48)(-12.000,-17.478){2}{\rule{0.400pt}{0.367pt}}
\multiput(846.92,1010.16)(-0.493,-0.734){23}{\rule{0.119pt}{0.685pt}}
\multiput(847.17,1011.58)(-13.000,-17.579){2}{\rule{0.400pt}{0.342pt}}
\multiput(833.92,991.09)(-0.492,-0.755){21}{\rule{0.119pt}{0.700pt}}
\multiput(834.17,992.55)(-12.000,-16.547){2}{\rule{0.400pt}{0.350pt}}
\multiput(821.92,973.29)(-0.493,-0.695){23}{\rule{0.119pt}{0.654pt}}
\multiput(822.17,974.64)(-13.000,-16.643){2}{\rule{0.400pt}{0.327pt}}
\multiput(808.92,955.29)(-0.493,-0.695){23}{\rule{0.119pt}{0.654pt}}
\multiput(809.17,956.64)(-13.000,-16.643){2}{\rule{0.400pt}{0.327pt}}
\multiput(795.92,936.96)(-0.492,-0.798){21}{\rule{0.119pt}{0.733pt}}
\multiput(796.17,938.48)(-12.000,-17.478){2}{\rule{0.400pt}{0.367pt}}
\multiput(783.92,912.03)(-0.493,-2.638){23}{\rule{0.119pt}{2.162pt}}
\multiput(784.17,916.51)(-13.000,-62.514){2}{\rule{0.400pt}{1.081pt}}
\multiput(770.92,850.40)(-0.492,-0.970){21}{\rule{0.119pt}{0.867pt}}
\multiput(771.17,852.20)(-12.000,-21.201){2}{\rule{0.400pt}{0.433pt}}
\multiput(758.92,827.90)(-0.493,-0.814){23}{\rule{0.119pt}{0.746pt}}
\multiput(759.17,829.45)(-13.000,-19.451){2}{\rule{0.400pt}{0.373pt}}
\multiput(745.92,806.26)(-0.492,-1.013){21}{\rule{0.119pt}{0.900pt}}
\multiput(746.17,808.13)(-12.000,-22.132){2}{\rule{0.400pt}{0.450pt}}
\multiput(733.92,781.24)(-0.493,-1.329){23}{\rule{0.119pt}{1.146pt}}
\multiput(734.17,783.62)(-13.000,-31.621){2}{\rule{0.400pt}{0.573pt}}
\multiput(720.92,747.37)(-0.493,-1.290){23}{\rule{0.119pt}{1.115pt}}
\multiput(721.17,749.68)(-13.000,-30.685){2}{\rule{0.400pt}{0.558pt}}
\multiput(707.92,715.40)(-0.492,-0.970){21}{\rule{0.119pt}{0.867pt}}
\multiput(708.17,717.20)(-12.000,-21.201){2}{\rule{0.400pt}{0.433pt}}
\multiput(695.92,693.16)(-0.493,-0.734){23}{\rule{0.119pt}{0.685pt}}
\multiput(696.17,694.58)(-13.000,-17.579){2}{\rule{0.400pt}{0.342pt}}
\multiput(682.92,673.82)(-0.492,-0.841){21}{\rule{0.119pt}{0.767pt}}
\multiput(683.17,675.41)(-12.000,-18.409){2}{\rule{0.400pt}{0.383pt}}
\multiput(670.92,654.41)(-0.493,-0.655){23}{\rule{0.119pt}{0.623pt}}
\multiput(671.17,655.71)(-13.000,-15.707){2}{\rule{0.400pt}{0.312pt}}
\multiput(657.92,637.29)(-0.493,-0.695){23}{\rule{0.119pt}{0.654pt}}
\multiput(658.17,638.64)(-13.000,-16.643){2}{\rule{0.400pt}{0.327pt}}
\multiput(644.92,619.37)(-0.492,-0.669){21}{\rule{0.119pt}{0.633pt}}
\multiput(645.17,620.69)(-12.000,-14.685){2}{\rule{0.400pt}{0.317pt}}
\multiput(632.92,603.41)(-0.493,-0.655){23}{\rule{0.119pt}{0.623pt}}
\multiput(633.17,604.71)(-13.000,-15.707){2}{\rule{0.400pt}{0.312pt}}
\multiput(619.92,586.51)(-0.492,-0.625){21}{\rule{0.119pt}{0.600pt}}
\multiput(620.17,587.75)(-12.000,-13.755){2}{\rule{0.400pt}{0.300pt}}
\multiput(607.92,571.54)(-0.493,-0.616){23}{\rule{0.119pt}{0.592pt}}
\multiput(608.17,572.77)(-13.000,-14.771){2}{\rule{0.400pt}{0.296pt}}
\multiput(594.92,555.51)(-0.492,-0.625){21}{\rule{0.119pt}{0.600pt}}
\multiput(595.17,556.75)(-12.000,-13.755){2}{\rule{0.400pt}{0.300pt}}
\multiput(582.92,540.80)(-0.493,-0.536){23}{\rule{0.119pt}{0.531pt}}
\multiput(583.17,541.90)(-13.000,-12.898){2}{\rule{0.400pt}{0.265pt}}
\multiput(569.92,526.67)(-0.493,-0.576){23}{\rule{0.119pt}{0.562pt}}
\multiput(570.17,527.83)(-13.000,-13.834){2}{\rule{0.400pt}{0.281pt}}
\multiput(556.92,511.65)(-0.492,-0.582){21}{\rule{0.119pt}{0.567pt}}
\multiput(557.17,512.82)(-12.000,-12.824){2}{\rule{0.400pt}{0.283pt}}
\multiput(544.92,497.67)(-0.493,-0.576){23}{\rule{0.119pt}{0.562pt}}
\multiput(545.17,498.83)(-13.000,-13.834){2}{\rule{0.400pt}{0.281pt}}
\multiput(531.92,482.65)(-0.492,-0.582){21}{\rule{0.119pt}{0.567pt}}
\multiput(532.17,483.82)(-12.000,-12.824){2}{\rule{0.400pt}{0.283pt}}
\multiput(519.92,468.80)(-0.493,-0.536){23}{\rule{0.119pt}{0.531pt}}
\multiput(520.17,469.90)(-13.000,-12.898){2}{\rule{0.400pt}{0.265pt}}
\multiput(506.92,454.79)(-0.492,-0.539){21}{\rule{0.119pt}{0.533pt}}
\multiput(507.17,455.89)(-12.000,-11.893){2}{\rule{0.400pt}{0.267pt}}
\multiput(494.92,441.80)(-0.493,-0.536){23}{\rule{0.119pt}{0.531pt}}
\multiput(495.17,442.90)(-13.000,-12.898){2}{\rule{0.400pt}{0.265pt}}
\multiput(481.92,427.80)(-0.493,-0.536){23}{\rule{0.119pt}{0.531pt}}
\multiput(482.17,428.90)(-13.000,-12.898){2}{\rule{0.400pt}{0.265pt}}
\multiput(468.92,413.79)(-0.492,-0.539){21}{\rule{0.119pt}{0.533pt}}
\multiput(469.17,414.89)(-12.000,-11.893){2}{\rule{0.400pt}{0.267pt}}
\multiput(455.92,401.92)(-0.497,-0.493){23}{\rule{0.500pt}{0.119pt}}
\multiput(456.96,402.17)(-11.962,-13.000){2}{\rule{0.250pt}{0.400pt}}
\multiput(443.92,387.79)(-0.492,-0.539){21}{\rule{0.119pt}{0.533pt}}
\multiput(444.17,388.89)(-12.000,-11.893){2}{\rule{0.400pt}{0.267pt}}
\multiput(431.92,374.80)(-0.493,-0.536){23}{\rule{0.119pt}{0.531pt}}
\multiput(432.17,375.90)(-13.000,-12.898){2}{\rule{0.400pt}{0.265pt}}
\multiput(417.92,361.92)(-0.497,-0.493){23}{\rule{0.500pt}{0.119pt}}
\multiput(418.96,362.17)(-11.962,-13.000){2}{\rule{0.250pt}{0.400pt}}
\multiput(405.92,347.79)(-0.492,-0.539){21}{\rule{0.119pt}{0.533pt}}
\multiput(406.17,348.89)(-12.000,-11.893){2}{\rule{0.400pt}{0.267pt}}
\multiput(392.79,335.92)(-0.539,-0.492){21}{\rule{0.533pt}{0.119pt}}
\multiput(393.89,336.17)(-11.893,-12.000){2}{\rule{0.267pt}{0.400pt}}
\multiput(380.92,322.79)(-0.492,-0.539){21}{\rule{0.119pt}{0.533pt}}
\multiput(381.17,323.89)(-12.000,-11.893){2}{\rule{0.400pt}{0.267pt}}
\multiput(367.92,310.92)(-0.497,-0.493){23}{\rule{0.500pt}{0.119pt}}
\multiput(368.96,311.17)(-11.962,-13.000){2}{\rule{0.250pt}{0.400pt}}
\multiput(355.92,296.79)(-0.492,-0.539){21}{\rule{0.119pt}{0.533pt}}
\multiput(356.17,297.89)(-12.000,-11.893){2}{\rule{0.400pt}{0.267pt}}
\multiput(342.79,284.92)(-0.539,-0.492){21}{\rule{0.533pt}{0.119pt}}
\multiput(343.89,285.17)(-11.893,-12.000){2}{\rule{0.267pt}{0.400pt}}
\multiput(329.92,272.92)(-0.497,-0.493){23}{\rule{0.500pt}{0.119pt}}
\multiput(330.96,273.17)(-11.962,-13.000){2}{\rule{0.250pt}{0.400pt}}
\multiput(316.92,259.92)(-0.496,-0.492){21}{\rule{0.500pt}{0.119pt}}
\multiput(317.96,260.17)(-10.962,-12.000){2}{\rule{0.250pt}{0.400pt}}
\multiput(304.92,247.92)(-0.497,-0.493){23}{\rule{0.500pt}{0.119pt}}
\multiput(305.96,248.17)(-11.962,-13.000){2}{\rule{0.250pt}{0.400pt}}
\multiput(291.92,234.92)(-0.496,-0.492){21}{\rule{0.500pt}{0.119pt}}
\multiput(292.96,235.17)(-10.962,-12.000){2}{\rule{0.250pt}{0.400pt}}
\sbox{\plotpoint}{\rule[-0.400pt]{0.800pt}{0.800pt}}%
\put(948,1371){\vector(0,-1){87}}
\put(772,645){\vector(0,1){117}}
\put(722,471){\vector(0,1){116}}
\end{picture}

\begin{center}{\Large{\bf Fig.1(b)}}\end{center}

\noindent {\bf Fig.1(b):} The temperature variation of the dynamic energy ($E$) for $D = 0.2$, $h_{0x} = 0.3$ and $h_{0z} = 1.0$. 
The vertical arrows represent the transition points.
\newpage
\setlength{\unitlength}{0.240900pt}
\ifx\plotpoint\undefined\newsavebox{\plotpoint}\fi
\sbox{\plotpoint}{\rule[-0.200pt]{0.400pt}{0.400pt}}%
\begin{picture}(1500,1440)(0,0)
\font\gnuplot=cmr10 at 10pt
\gnuplot
\sbox{\plotpoint}{\rule[-0.200pt]{0.400pt}{0.400pt}}%
\put(121.0,123.0){\rule[-0.200pt]{4.818pt}{0.400pt}}
\put(101,123){\makebox(0,0)[r]{0}}
\put(1419.0,123.0){\rule[-0.200pt]{4.818pt}{0.400pt}}
\put(121.0,378.0){\rule[-0.200pt]{4.818pt}{0.400pt}}
\put(101,378){\makebox(0,0)[r]{1}}
\put(1419.0,378.0){\rule[-0.200pt]{4.818pt}{0.400pt}}
\put(121.0,634.0){\rule[-0.200pt]{4.818pt}{0.400pt}}
\put(101,634){\makebox(0,0)[r]{2}}
\put(1419.0,634.0){\rule[-0.200pt]{4.818pt}{0.400pt}}
\put(121.0,889.0){\rule[-0.200pt]{4.818pt}{0.400pt}}
\put(101,889){\makebox(0,0)[r]{3}}
\put(1419.0,889.0){\rule[-0.200pt]{4.818pt}{0.400pt}}
\put(121.0,1145.0){\rule[-0.200pt]{4.818pt}{0.400pt}}
\put(101,1145){\makebox(0,0)[r]{4}}
\put(1419.0,1145.0){\rule[-0.200pt]{4.818pt}{0.400pt}}
\put(121.0,1400.0){\rule[-0.200pt]{4.818pt}{0.400pt}}
\put(101,1400){\makebox(0,0)[r]{5}}
\put(1419.0,1400.0){\rule[-0.200pt]{4.818pt}{0.400pt}}
\put(121.0,123.0){\rule[-0.200pt]{0.400pt}{4.818pt}}
\put(121,82){\makebox(0,0){0}}
\put(121.0,1380.0){\rule[-0.200pt]{0.400pt}{4.818pt}}
\put(451.0,123.0){\rule[-0.200pt]{0.400pt}{4.818pt}}
\put(451,82){\makebox(0,0){0.5}}
\put(451.0,1380.0){\rule[-0.200pt]{0.400pt}{4.818pt}}
\put(780.0,123.0){\rule[-0.200pt]{0.400pt}{4.818pt}}
\put(780,82){\makebox(0,0){1}}
\put(780.0,1380.0){\rule[-0.200pt]{0.400pt}{4.818pt}}
\put(1110.0,123.0){\rule[-0.200pt]{0.400pt}{4.818pt}}
\put(1110,82){\makebox(0,0){1.5}}
\put(1110.0,1380.0){\rule[-0.200pt]{0.400pt}{4.818pt}}
\put(1439.0,123.0){\rule[-0.200pt]{0.400pt}{4.818pt}}
\put(1439,82){\makebox(0,0){2}}
\put(1439.0,1380.0){\rule[-0.200pt]{0.400pt}{4.818pt}}
\put(121.0,123.0){\rule[-0.200pt]{317.506pt}{0.400pt}}
\put(1439.0,123.0){\rule[-0.200pt]{0.400pt}{307.629pt}}
\put(121.0,1400.0){\rule[-0.200pt]{317.506pt}{0.400pt}}
\put(40,761){\makebox(0,0){$C$}}
\put(780,21){\makebox(0,0){$T$}}
\put(1373,1300){\makebox(0,0){\Large {\bf (c)}}}
\put(1110,1145){\makebox(0,0)[l]{$D=0.2$}}
\put(1110,1017){\makebox(0,0)[l]{$h_{0x} = 0.3$}}
\put(1110,889){\makebox(0,0)[l]{$h_{0z} = 1.0$}}
\put(121.0,123.0){\rule[-0.200pt]{0.400pt}{307.629pt}}
\sbox{\plotpoint}{\rule[-0.500pt]{1.000pt}{1.000pt}}%
\put(1294,229){\circle*{24}}
\put(1281,221){\circle*{24}}
\put(1268,218){\circle*{24}}
\put(1254,221){\circle*{24}}
\put(1241,216){\circle*{24}}
\put(1228,209){\circle*{24}}
\put(1215,208){\circle*{24}}
\put(1202,205){\circle*{24}}
\put(1189,200){\circle*{24}}
\put(1175,196){\circle*{24}}
\put(1162,196){\circle*{24}}
\put(1149,190){\circle*{24}}
\put(1136,186){\circle*{24}}
\put(1123,187){\circle*{24}}
\put(1110,187){\circle*{24}}
\put(1096,181){\circle*{24}}
\put(1083,178){\circle*{24}}
\put(1070,185){\circle*{24}}
\put(1057,182){\circle*{24}}
\put(1044,182){\circle*{24}}
\put(1030,196){\circle*{24}}
\put(1017,203){\circle*{24}}
\put(1004,213){\circle*{24}}
\put(991,240){\circle*{24}}
\put(978,262){\circle*{24}}
\put(965,318){\circle*{24}}
\put(951,482){\circle*{24}}
\put(938,667){\circle*{24}}
\put(925,716){\circle*{24}}
\put(912,662){\circle*{24}}
\put(899,624){\circle*{24}}
\put(885,613){\circle*{24}}
\put(872,585){\circle*{24}}
\put(859,566){\circle*{24}}
\put(846,562){\circle*{24}}
\put(833,548){\circle*{24}}
\put(820,539){\circle*{24}}
\put(806,530){\circle*{24}}
\put(793,527){\circle*{24}}
\put(780,523){\circle*{24}}
\put(767,530){\circle*{24}}
\put(754,1068){\circle*{24}}
\put(740,1108){\circle*{24}}
\put(727,609){\circle*{24}}
\put(714,618){\circle*{24}}
\put(701,758){\circle*{24}}
\put(688,857){\circle*{24}}
\put(675,737){\circle*{24}}
\put(661,593){\circle*{24}}
\put(648,551){\circle*{24}}
\put(635,528){\circle*{24}}
\put(622,507){\circle*{24}}
\put(609,499){\circle*{24}}
\put(595,485){\circle*{24}}
\put(582,469){\circle*{24}}
\put(569,466){\circle*{24}}
\put(556,461){\circle*{24}}
\put(543,445){\circle*{24}}
\put(530,448){\circle*{24}}
\put(516,444){\circle*{24}}
\put(503,437){\circle*{24}}
\put(490,437){\circle*{24}}
\put(477,430){\circle*{24}}
\put(464,426){\circle*{24}}
\put(451,425){\circle*{24}}
\put(437,423){\circle*{24}}
\put(424,419){\circle*{24}}
\put(411,417){\circle*{24}}
\put(398,413){\circle*{24}}
\put(385,413){\circle*{24}}
\put(371,413){\circle*{24}}
\put(358,408){\circle*{24}}
\put(345,406){\circle*{24}}
\put(332,406){\circle*{24}}
\put(319,405){\circle*{24}}
\put(306,402){\circle*{24}}
\put(292,399){\circle*{24}}
\put(279,400){\circle*{24}}
\put(266,399){\circle*{24}}
\put(253,396){\circle*{24}}
\put(240,396){\circle*{24}}
\sbox{\plotpoint}{\rule[-0.200pt]{0.400pt}{0.400pt}}%
\put(1294,229){\usebox{\plotpoint}}
\multiput(1290.89,227.93)(-0.824,-0.488){13}{\rule{0.750pt}{0.117pt}}
\multiput(1292.44,228.17)(-11.443,-8.000){2}{\rule{0.375pt}{0.400pt}}
\multiput(1273.39,219.95)(-2.695,-0.447){3}{\rule{1.833pt}{0.108pt}}
\multiput(1277.19,220.17)(-9.195,-3.000){2}{\rule{0.917pt}{0.400pt}}
\multiput(1259.84,218.61)(-2.918,0.447){3}{\rule{1.967pt}{0.108pt}}
\multiput(1263.92,217.17)(-9.918,3.000){2}{\rule{0.983pt}{0.400pt}}
\multiput(1249.27,219.93)(-1.378,-0.477){7}{\rule{1.140pt}{0.115pt}}
\multiput(1251.63,220.17)(-10.634,-5.000){2}{\rule{0.570pt}{0.400pt}}
\multiput(1237.50,214.93)(-0.950,-0.485){11}{\rule{0.843pt}{0.117pt}}
\multiput(1239.25,215.17)(-11.251,-7.000){2}{\rule{0.421pt}{0.400pt}}
\put(1215,207.67){\rule{3.132pt}{0.400pt}}
\multiput(1221.50,208.17)(-6.500,-1.000){2}{\rule{1.566pt}{0.400pt}}
\multiput(1207.39,206.95)(-2.695,-0.447){3}{\rule{1.833pt}{0.108pt}}
\multiput(1211.19,207.17)(-9.195,-3.000){2}{\rule{0.917pt}{0.400pt}}
\multiput(1197.27,203.93)(-1.378,-0.477){7}{\rule{1.140pt}{0.115pt}}
\multiput(1199.63,204.17)(-10.634,-5.000){2}{\rule{0.570pt}{0.400pt}}
\multiput(1182.77,198.94)(-1.943,-0.468){5}{\rule{1.500pt}{0.113pt}}
\multiput(1185.89,199.17)(-10.887,-4.000){2}{\rule{0.750pt}{0.400pt}}
\multiput(1157.99,194.93)(-1.123,-0.482){9}{\rule{0.967pt}{0.116pt}}
\multiput(1159.99,195.17)(-10.994,-6.000){2}{\rule{0.483pt}{0.400pt}}
\multiput(1143.19,188.94)(-1.797,-0.468){5}{\rule{1.400pt}{0.113pt}}
\multiput(1146.09,189.17)(-10.094,-4.000){2}{\rule{0.700pt}{0.400pt}}
\put(1123,185.67){\rule{3.132pt}{0.400pt}}
\multiput(1129.50,185.17)(-6.500,1.000){2}{\rule{1.566pt}{0.400pt}}
\put(1162.0,196.0){\rule[-0.200pt]{3.132pt}{0.400pt}}
\multiput(1105.71,185.93)(-1.214,-0.482){9}{\rule{1.033pt}{0.116pt}}
\multiput(1107.86,186.17)(-11.855,-6.000){2}{\rule{0.517pt}{0.400pt}}
\multiput(1088.39,179.95)(-2.695,-0.447){3}{\rule{1.833pt}{0.108pt}}
\multiput(1092.19,180.17)(-9.195,-3.000){2}{\rule{0.917pt}{0.400pt}}
\multiput(1079.50,178.59)(-0.950,0.485){11}{\rule{0.843pt}{0.117pt}}
\multiput(1081.25,177.17)(-11.251,7.000){2}{\rule{0.421pt}{0.400pt}}
\multiput(1062.39,183.95)(-2.695,-0.447){3}{\rule{1.833pt}{0.108pt}}
\multiput(1066.19,184.17)(-9.195,-3.000){2}{\rule{0.917pt}{0.400pt}}
\put(1110.0,187.0){\rule[-0.200pt]{3.132pt}{0.400pt}}
\multiput(1041.92,182.58)(-0.497,0.494){25}{\rule{0.500pt}{0.119pt}}
\multiput(1042.96,181.17)(-12.962,14.000){2}{\rule{0.250pt}{0.400pt}}
\multiput(1026.50,196.59)(-0.950,0.485){11}{\rule{0.843pt}{0.117pt}}
\multiput(1028.25,195.17)(-11.251,7.000){2}{\rule{0.421pt}{0.400pt}}
\multiput(1014.43,203.58)(-0.652,0.491){17}{\rule{0.620pt}{0.118pt}}
\multiput(1015.71,202.17)(-11.713,10.000){2}{\rule{0.310pt}{0.400pt}}
\multiput(1002.92,213.00)(-0.493,1.052){23}{\rule{0.119pt}{0.931pt}}
\multiput(1003.17,213.00)(-13.000,25.068){2}{\rule{0.400pt}{0.465pt}}
\multiput(989.92,240.00)(-0.493,0.853){23}{\rule{0.119pt}{0.777pt}}
\multiput(990.17,240.00)(-13.000,20.387){2}{\rule{0.400pt}{0.388pt}}
\multiput(976.92,262.00)(-0.493,2.201){23}{\rule{0.119pt}{1.823pt}}
\multiput(977.17,262.00)(-13.000,52.216){2}{\rule{0.400pt}{0.912pt}}
\multiput(963.92,318.00)(-0.494,6.006){25}{\rule{0.119pt}{4.786pt}}
\multiput(964.17,318.00)(-14.000,154.067){2}{\rule{0.400pt}{2.393pt}}
\multiput(949.92,482.00)(-0.493,7.316){23}{\rule{0.119pt}{5.792pt}}
\multiput(950.17,482.00)(-13.000,172.978){2}{\rule{0.400pt}{2.896pt}}
\multiput(936.92,667.00)(-0.493,1.924){23}{\rule{0.119pt}{1.608pt}}
\multiput(937.17,667.00)(-13.000,45.663){2}{\rule{0.400pt}{0.804pt}}
\multiput(923.92,708.69)(-0.493,-2.122){23}{\rule{0.119pt}{1.762pt}}
\multiput(924.17,712.34)(-13.000,-50.344){2}{\rule{0.400pt}{0.881pt}}
\multiput(910.92,656.73)(-0.493,-1.488){23}{\rule{0.119pt}{1.269pt}}
\multiput(911.17,659.37)(-13.000,-35.366){2}{\rule{0.400pt}{0.635pt}}
\multiput(896.47,622.92)(-0.637,-0.492){19}{\rule{0.609pt}{0.118pt}}
\multiput(897.74,623.17)(-12.736,-11.000){2}{\rule{0.305pt}{0.400pt}}
\multiput(883.92,609.01)(-0.493,-1.091){23}{\rule{0.119pt}{0.962pt}}
\multiput(884.17,611.00)(-13.000,-26.004){2}{\rule{0.400pt}{0.481pt}}
\multiput(870.92,582.16)(-0.493,-0.734){23}{\rule{0.119pt}{0.685pt}}
\multiput(871.17,583.58)(-13.000,-17.579){2}{\rule{0.400pt}{0.342pt}}
\multiput(853.19,564.94)(-1.797,-0.468){5}{\rule{1.400pt}{0.113pt}}
\multiput(856.09,565.17)(-10.094,-4.000){2}{\rule{0.700pt}{0.400pt}}
\multiput(844.92,559.80)(-0.493,-0.536){23}{\rule{0.119pt}{0.531pt}}
\multiput(845.17,560.90)(-13.000,-12.898){2}{\rule{0.400pt}{0.265pt}}
\multiput(830.19,546.93)(-0.728,-0.489){15}{\rule{0.678pt}{0.118pt}}
\multiput(831.59,547.17)(-11.593,-9.000){2}{\rule{0.339pt}{0.400pt}}
\multiput(817.00,537.93)(-0.786,-0.489){15}{\rule{0.722pt}{0.118pt}}
\multiput(818.50,538.17)(-12.501,-9.000){2}{\rule{0.361pt}{0.400pt}}
\multiput(798.39,528.95)(-2.695,-0.447){3}{\rule{1.833pt}{0.108pt}}
\multiput(802.19,529.17)(-9.195,-3.000){2}{\rule{0.917pt}{0.400pt}}
\multiput(787.19,525.94)(-1.797,-0.468){5}{\rule{1.400pt}{0.113pt}}
\multiput(790.09,526.17)(-10.094,-4.000){2}{\rule{0.700pt}{0.400pt}}
\multiput(776.50,523.59)(-0.950,0.485){11}{\rule{0.843pt}{0.117pt}}
\multiput(778.25,522.17)(-11.251,7.000){2}{\rule{0.421pt}{0.400pt}}
\multiput(765.92,530.00)(-0.493,21.312){23}{\rule{0.119pt}{16.654pt}}
\multiput(766.17,530.00)(-13.000,503.434){2}{\rule{0.400pt}{8.327pt}}
\multiput(752.92,1068.00)(-0.494,1.452){25}{\rule{0.119pt}{1.243pt}}
\multiput(753.17,1068.00)(-14.000,37.420){2}{\rule{0.400pt}{0.621pt}}
\multiput(738.92,1043.85)(-0.493,-19.766){23}{\rule{0.119pt}{15.454pt}}
\multiput(739.17,1075.92)(-13.000,-466.925){2}{\rule{0.400pt}{7.727pt}}
\multiput(724.19,609.59)(-0.728,0.489){15}{\rule{0.678pt}{0.118pt}}
\multiput(725.59,608.17)(-11.593,9.000){2}{\rule{0.339pt}{0.400pt}}
\multiput(712.92,618.00)(-0.493,5.532){23}{\rule{0.119pt}{4.408pt}}
\multiput(713.17,618.00)(-13.000,130.852){2}{\rule{0.400pt}{2.204pt}}
\multiput(699.92,758.00)(-0.493,3.906){23}{\rule{0.119pt}{3.146pt}}
\multiput(700.17,758.00)(-13.000,92.470){2}{\rule{0.400pt}{1.573pt}}
\multiput(686.92,841.26)(-0.493,-4.739){23}{\rule{0.119pt}{3.792pt}}
\multiput(687.17,849.13)(-13.000,-112.129){2}{\rule{0.400pt}{1.896pt}}
\multiput(673.92,719.51)(-0.494,-5.271){25}{\rule{0.119pt}{4.214pt}}
\multiput(674.17,728.25)(-14.000,-135.253){2}{\rule{0.400pt}{2.107pt}}
\multiput(659.92,587.22)(-0.493,-1.646){23}{\rule{0.119pt}{1.392pt}}
\multiput(660.17,590.11)(-13.000,-39.110){2}{\rule{0.400pt}{0.696pt}}
\multiput(646.92,547.65)(-0.493,-0.893){23}{\rule{0.119pt}{0.808pt}}
\multiput(647.17,549.32)(-13.000,-21.324){2}{\rule{0.400pt}{0.404pt}}
\multiput(633.92,524.90)(-0.493,-0.814){23}{\rule{0.119pt}{0.746pt}}
\multiput(634.17,526.45)(-13.000,-19.451){2}{\rule{0.400pt}{0.373pt}}
\multiput(618.89,505.93)(-0.824,-0.488){13}{\rule{0.750pt}{0.117pt}}
\multiput(620.44,506.17)(-11.443,-8.000){2}{\rule{0.375pt}{0.400pt}}
\multiput(606.92,497.92)(-0.497,-0.494){25}{\rule{0.500pt}{0.119pt}}
\multiput(607.96,498.17)(-12.962,-14.000){2}{\rule{0.250pt}{0.400pt}}
\multiput(593.92,482.54)(-0.493,-0.616){23}{\rule{0.119pt}{0.592pt}}
\multiput(594.17,483.77)(-13.000,-14.771){2}{\rule{0.400pt}{0.296pt}}
\multiput(574.39,467.95)(-2.695,-0.447){3}{\rule{1.833pt}{0.108pt}}
\multiput(578.19,468.17)(-9.195,-3.000){2}{\rule{0.917pt}{0.400pt}}
\multiput(564.27,464.93)(-1.378,-0.477){7}{\rule{1.140pt}{0.115pt}}
\multiput(566.63,465.17)(-10.634,-5.000){2}{\rule{0.570pt}{0.400pt}}
\multiput(554.92,458.54)(-0.493,-0.616){23}{\rule{0.119pt}{0.592pt}}
\multiput(555.17,459.77)(-13.000,-14.771){2}{\rule{0.400pt}{0.296pt}}
\multiput(535.39,445.61)(-2.695,0.447){3}{\rule{1.833pt}{0.108pt}}
\multiput(539.19,444.17)(-9.195,3.000){2}{\rule{0.917pt}{0.400pt}}
\multiput(523.77,446.94)(-1.943,-0.468){5}{\rule{1.500pt}{0.113pt}}
\multiput(526.89,447.17)(-10.887,-4.000){2}{\rule{0.750pt}{0.400pt}}
\multiput(512.50,442.93)(-0.950,-0.485){11}{\rule{0.843pt}{0.117pt}}
\multiput(514.25,443.17)(-11.251,-7.000){2}{\rule{0.421pt}{0.400pt}}
\put(1044.0,182.0){\rule[-0.200pt]{3.132pt}{0.400pt}}
\multiput(486.50,435.93)(-0.950,-0.485){11}{\rule{0.843pt}{0.117pt}}
\multiput(488.25,436.17)(-11.251,-7.000){2}{\rule{0.421pt}{0.400pt}}
\multiput(471.19,428.94)(-1.797,-0.468){5}{\rule{1.400pt}{0.113pt}}
\multiput(474.09,429.17)(-10.094,-4.000){2}{\rule{0.700pt}{0.400pt}}
\put(451,424.67){\rule{3.132pt}{0.400pt}}
\multiput(457.50,425.17)(-6.500,-1.000){2}{\rule{1.566pt}{0.400pt}}
\put(437,423.17){\rule{2.900pt}{0.400pt}}
\multiput(444.98,424.17)(-7.981,-2.000){2}{\rule{1.450pt}{0.400pt}}
\multiput(431.19,421.94)(-1.797,-0.468){5}{\rule{1.400pt}{0.113pt}}
\multiput(434.09,422.17)(-10.094,-4.000){2}{\rule{0.700pt}{0.400pt}}
\put(411,417.17){\rule{2.700pt}{0.400pt}}
\multiput(418.40,418.17)(-7.396,-2.000){2}{\rule{1.350pt}{0.400pt}}
\multiput(405.19,415.94)(-1.797,-0.468){5}{\rule{1.400pt}{0.113pt}}
\multiput(408.09,416.17)(-10.094,-4.000){2}{\rule{0.700pt}{0.400pt}}
\put(490.0,437.0){\rule[-0.200pt]{3.132pt}{0.400pt}}
\multiput(366.27,411.93)(-1.378,-0.477){7}{\rule{1.140pt}{0.115pt}}
\multiput(368.63,412.17)(-10.634,-5.000){2}{\rule{0.570pt}{0.400pt}}
\put(345,406.17){\rule{2.700pt}{0.400pt}}
\multiput(352.40,407.17)(-7.396,-2.000){2}{\rule{1.350pt}{0.400pt}}
\put(371.0,413.0){\rule[-0.200pt]{6.504pt}{0.400pt}}
\put(319,404.67){\rule{3.132pt}{0.400pt}}
\multiput(325.50,405.17)(-6.500,-1.000){2}{\rule{1.566pt}{0.400pt}}
\multiput(311.39,403.95)(-2.695,-0.447){3}{\rule{1.833pt}{0.108pt}}
\multiput(315.19,404.17)(-9.195,-3.000){2}{\rule{0.917pt}{0.400pt}}
\multiput(297.84,400.95)(-2.918,-0.447){3}{\rule{1.967pt}{0.108pt}}
\multiput(301.92,401.17)(-9.918,-3.000){2}{\rule{0.983pt}{0.400pt}}
\put(279,398.67){\rule{3.132pt}{0.400pt}}
\multiput(285.50,398.17)(-6.500,1.000){2}{\rule{1.566pt}{0.400pt}}
\put(266,398.67){\rule{3.132pt}{0.400pt}}
\multiput(272.50,399.17)(-6.500,-1.000){2}{\rule{1.566pt}{0.400pt}}
\multiput(258.39,397.95)(-2.695,-0.447){3}{\rule{1.833pt}{0.108pt}}
\multiput(262.19,398.17)(-9.195,-3.000){2}{\rule{0.917pt}{0.400pt}}
\put(332.0,406.0){\rule[-0.200pt]{3.132pt}{0.400pt}}
\put(240.0,396.0){\rule[-0.200pt]{3.132pt}{0.400pt}}
\sbox{\plotpoint}{\rule[-0.400pt]{0.800pt}{0.800pt}}%
\put(925,889){\vector(0,-1){101}}
\put(740,1272){\vector(0,-1){101}}
\put(688,1017){\vector(0,-1){101}}
\end{picture}

\begin{center}{\Large{\bf Fig.1(c)}}\end{center}

\noindent {\bf Fig.1(c):} The temperature variation of dynamic specific heat ($C = {{dE} \over {dT}}$) for $D = 0.2$, $h_{0x} = 0.3$
and $h_{0z} = 1.0$. Vertical arrows show the peaks and the transition points.
\end{document}